\documentclass[a4paper,twocolumn,11pt,accepted=2024-11-26]{quantumarticle}
\pdfoutput=1
\usepackage[utf8]{inputenc}
\usepackage[english]{babel}
\usepackage[T1]{fontenc}
\usepackage[numbers]{natbib}

\usepackage{graphicx}
\usepackage{latexsym}
\usepackage{amsmath,amssymb}
\usepackage{amsfonts}
\usepackage{upgreek}
\usepackage{bm}
\usepackage{multirow}
\usepackage{enumitem}
\usepackage{color}
\usepackage{hyperref}
\usepackage{physics}
\usepackage{tcolorbox}
\usepackage{manfnt}
\usepackage{relsize}

\newcommand{\beq}{\begin{equation}}
\newcommand{\eeq}{\end{equation}}
\newcommand{\beqn}{\begin{eqnarray}}
\newcommand{\eeqn}{\end{eqnarray}}

\newcommand{\ra}{\rightarrow}

\newcommand{\cA}{ {\cal A} }

\newcommand{\cD}{ {\cal D} }
\newcommand{\cE}{ {\cal E} }

\newcommand{\cH}{ {\cal H} }

\newcommand{\cL}{ {\cal L} }
\newcommand{\cP}{ {\cal P} }
\newcommand{\cS}{ {\cal S} }

\newcommand{\ii}{\mathrm{i}}

\newcommand{\be}{\mathbf{e}}

\newcommand{\SO}{\mathrm{SO}}
\renewcommand{\O}{\mathrm{O}}

\newcommand{\U}{\mathrm{U}}

\newcommand{\vect}[1]{{\bm{#1}}}

\newcommand{\bsigma}{{\bm{\sigma}}}

\newcommand{\bs}{{\bm{s}}}

\newcommand{\bmeta}{{\bm{\eta}}}
\newcommand{\secref}[1]{Sec.\,\ref{#1}}
\newcommand{\appref}[1]{Appendix.\,\ref{#1}}
\newcommand{\eqnref}[1]{Eq.\,\eqref{#1}}
\newcommand{\figref}[1]{Fig.\,\ref{#1}}

\newcommand{\rd}{\partial}

\newcommand{\vdagger}{{\vphantom{\dagger}}}

\newcommand{\Ket}[1]{{\Vert #1 \rangle\hspace{-2pt}\rangle}}
\newcommand{\Bra}[1]{{\langle\hspace{-2pt}\langle #1 \Vert}}

\newcommand{\spt}{\textrm{spt}}

\newcommand{\te}{{\textrm{e}}}
\newcommand{\oo}{{\textrm{o}}}
\newcommand{\teven}{{\textrm{even}}}
\newcommand{\todd}{{\textrm{odd}}}
\newcommand{\rAngle}{\rangle \hspace{-2pt} \rangle }
\newcommand{\lAngle}{\langle \hspace{-2pt} \langle }

\newcommand{\prb}{Phys. Rev. B}  
\newcommand{\prl}{Phys. Rev. Lett.}  
\newcommand{\prd}{Phys. Rev. D}  
\begin{document}

\title{Symmetry protected topological phases under decoherence}

\author{Jong Yeon Lee}

\affiliation{Kavli Institute for Theoretical Physics, University of California, Santa Barbara, CA 93106, USA}
\affiliation{Department of Physics, University of Illinois Urbana-Champaign, Urbana, IL 61801, USA}

\author{Yi-Zhuang You}

\affiliation{Department of Physics, University of California San
Diego, La Jolla, CA 92093, USA}

\author{Cenke Xu}

\affiliation{Department of Physics, University of California,
Santa Barbara, CA 93106, USA}

\begin{abstract}

We investigate mixed states exhibiting nontrivial topological features, focusing on symmetry-protected topological (SPT) phases under various types of decoherence. Our findings demonstrate that these systems can retain topological information from the SPT ground state despite decoherence. In the ''doubled Hilbert space,'' we define symmetry-protected topological ensembles (SPT ensembles) and examine boundary anomalies in this space. We generalize the concept of the strange correlator, initially used to diagnose SPT ground states, to identify anomalies in mixed-state density matrices. Through exact calculations of stabilizer Hamiltonians and field theory evaluations, we show that nontrivial features of SPT states persist in two types of strange correlators: type-I and type-II. The type-I strange correlator reveals SPT information that can be efficiently detected and used experimentally, such as in preparing long-range entangled states. The type-II strange correlator encodes the full topological response of the decohered mixed state, reflecting the SPT state's pre-decoherence presence. Our work offers a unified framework for understanding decohered SPT phases from an information-theoretic perspective.

\end{abstract}

\maketitle

\section{Introduction}

With rapid progress in various quantum simulator platforms, the study of quantum many-body states in noisy intermediate-scale quantum (NISQ) platforms has emerged as an active area of research, attracting growing interest from both the condensed matter and quantum information communities. In particular, various exotic quantum states have been realized in these platforms~\cite{scar2017, earlyToric, SPT2019, KZ2019, googleToric, SL2021, TimeCrystal}, demonstrating the possibility of realizing and manipulating quantum matter in near-term quantum devices. Among them, symmetry-protected topological (SPT) \cite{Gu0903.1069, Chen1008.3745, Pollmann0909.4059} phases are of great importance since the SPT states provide a class of non-trivially entangled quantum systems that can be realized in the NISQ platforms. Furthermore, they are relevant to important quantum information processing tasks, such as measurement-based quantum computation and state preparation~\cite{1Dcluster_GHZ, 2Dcluster, 2Dcluster_toric, 3dCluster_fracton1, 3dCluster_fracton2, Stephen2017, Raussendorf2019, NatMeasurement, NatRydberg,CSScode, CSScode2, ClusterCSS, Lu2022, Lee2022, Zhu2022, nonAbelianNat, nonAbelianNat2}. 
However, despite the fact that quantum decoherence of various types inevitably occurs in nature, most theoretical studies on information theoretic properties have been focusing on the {\it pure} SPT states.
Although various extensions of the notion of topological phases to mixed states have been explored in the past~\cite{statTI, statTI2, Kimchi1710.06860, Groot2021, MaWang2022, TO_thermal, TI_open1, TI_open2,TI_open3}, a general definition and diagnosis of symmetry protected topological ensembles (SPT ensemble) is still lacking and the discussions have not been extended to understand information-theoretic capacity of underlying topological states. 
Therefore, it is timely to ask what defining features of this highly interesting class of topological quantum phases can be identified under the presence of decoherence at a microscopic scale, a non-unitary perturbation prevalent in quantum devices. 

\begin{figure*}[!t]
    \centering
    \includegraphics[width = 0.99 \textwidth]{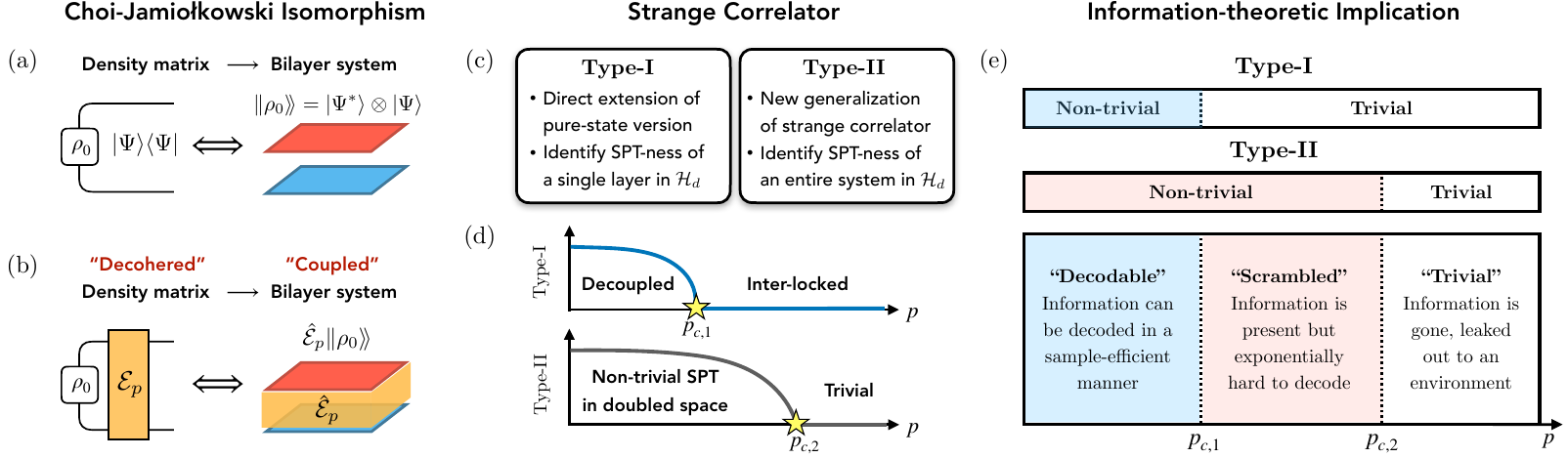}
    \caption{\label{fig:summary} {\bf Summary.} (a,b) Graphical representation of Choi-Jamiolkoski isomorphism, where pure (mixed) state density matrix maps to a decoupled (coupled) bilayer system living in a doubled Hilbert space. The channel $\cE_p$ whose decoherence strength is parametrized by $p$ would convert a pure state density matrix $\rho_0$ into a decohered density matrix $\cE_p[\rho_0]$. (c) In the doubled Hilbert space $\cH_d$, both types of strange correlators measure topological features (e.g. boundary ' t Hooft anomalies, topological responses) of a pure state $\hat{\cE}_p\Vert \rho_0 \rAngle$ in the doubled space: the type-I measures the topological features of a single layer, while the type-II measures the topological features of an entire doubled system. (d) Schematic behaviors of strange correlators as a function of the decoherence strength $p$, and their physical meanings in the doubled Hilbert space. (e) Schematic information-theoretic phase diagrams: nontrivial topological responses in the doubled Hilbert space have information-theoretic implications in the original Hilbert space. For $p < p_{c,1}$,  type-I SC is nontrivial and the SPT label information may be efficiently decoded (see \secref{sec:measure}). For $p_{c,1} < p < p_{c,2}$, type-II SC is nontrivial, and it corresponds to the information that the original state under decoherence was an SPT order. However, the quantity is very difficult to extract from the bulk measurements. For $p > p_{c,2}$, the SPT label information cannot be identified by the type-II strange correlator. Under the local decoherence models we considered in the lattice calculations, $p_{c,2}$ does not exist since the type-II strange correlator is always finite. However, for a more general decoherence model, we would expect the transition into the ``trivial'' information phase to exist.   }
\end{figure*}

It is commonly accepted that SPT states usually exhibit \emph{trivial} bulk features, in the sense that the properties such as the correlation between local order parameters are not distinguishable from those of trivial quantum disordered states. Since the defining symmetries of the SPT states are preserved, there are no bulk observables such as macroscopic magnetization. As a result, characteristic features of SPT states are often understood in terms of the nontrivial boundary physics, or nontrivial quantum numbers carried by the symmetry defects. However, it is not so easy to directly extract this information from a bulk ground state wave function of the SPT phase. In particular, when the defining symmetry of the SPT state involves spatial symmetries, an arbitrary open boundary may break the spatial symmetry and render the boundary state trivial unless the boundary is carefully designed to preserve all the spatial symmetries. In order to overcome these complexities, a notion \emph{strange correlator} was proposed~\cite{YouXu2013},  to diagnose nontrivial SPT properties using just the bulk ground state wave function, without reference to either boundaries or defects. With an analogy to the Wick-rotated correlator in the imaginary space-time at the interface against a trivial disordered state, the strange correlator defined purely based on the bulk wave function must be ``nontrivial'', $i.e.$ they must either saturate to a constant or decay as a power-law for $1d$ or $2d$ SPT states. As we will demonstrate in this work, the strange correlator can also be viewed as a general ``order parameter'' for the $1d$ and $2d$ SPT phases, analogous to the well-known string order parameter of the Haldane phase~\cite{SOP1989, SOP2008}. 
The strange correlator has been adopted as a useful tool in both conceptual understanding and numerical studies of SPT phases, for both bosonic and fermionic systems~\cite{scwierschem1,scwierschem2,scwierschem3,sczohar1,scmeng1,sczohar2,scmeng2,scwei,scmeng3,scwierschem4,sczhong,scscaffidi,scfrank1,scfrank2,scfrank3,scfrank4,scfrank5,schsieh,scfan,wu2020,scsagar,scmeng4}.

In experimental implementations, utilization, and verification of entangled quantum states, the states are subject to decoherence, which can erase the characteristic signature defined for a pure state. This key information of the original quantum state must be \emph{decoded} by the receiving end of the process. In this work, we demonstrate that SPT states under decoherence can retain nontrivial topological information that is in principle decodable, by exploiting the idea of strange correlators. One common decoherence mechanism in nature is thermalization, where at thermal equilibrium, all degrees of freedom of the system experience decoherence whose strength depends on the temperature. Such a ``massive'' thermal decoherence drives SPT phases with on-site symmetries to be trivial~\cite{SPT_finiteT_2017} This is the situation studied in various previous works such as Ref.~\cite{TI_open1,TI_open2,TI_open3}. Instead, we consider a ``selective'' decoherence to certain degrees of freedom of the system, which exposes an extremely rich possibility of physics under decoherence. With decoherence, the quantum state of interest becomes an \emph{ensemble} expressed by the density matrix. In the density matrix formalism, two types of strange correlators naturally emerge denoted as ``type-I'' and ``type-II'', whose graphical illustration is shown in \figref{fig:sc_pathintegral}(c,d). In the pure state limit, the former is essentially identical to the original definition of the strange correlator~\cite{YouXu2013}, while the latter is analogous to the Edwards-Anderson generalization of the original strange correlator.

From both exact lattice model and field-theoretic calculations, we will show that these two strange correlators serve as diagnoses for underlying decohered SPT states, providing a fundamental limit on one's information-theoretic capacity to process decohered SPT states (See~\figref{fig:information}).
In particular, nontrivial behaviors of type-I strange correlators correspond to one's capacity to efficiently probe and utilize decohered SPT states from experiments. Since the strange correlator is defined as the operator matrix element between two different states, it is experimentally challenging to measure and has been used mostly as a conceptual notion and a numerical tool to diagnose SPT phases in the past. However, we show that for a broad class of SPT states with decorated defect constructions, the \emph{marginalized} version of the type-I strange correlator can be experimentally probed using measurements and additional classical computational steps~\cite{Lee2022}. More interestingly, we reveal that this marginalized type-I strange correlator provides a unifying framework to understand the non-local order parameters of an SPT state and the feasibility of utilizing decohered SPT states, such as preparing a long-range entangled quantum state including topological states by measurement and decoding.

On the other hand, nontrivial behaviors of type-II strange correlators encode the full topological response of the mixed-state ensemble, and they correspond to one's fundamental capability to distinguish the decohered SPT state from trivial states. Therefore, the type-II strange correlator may serve as 
a general method of diagnosing mixed-state density matrices. Although the measurement of type-II strange correlators is expected to be more difficult than type-I strange correlators, they can be in principle probed by fidelity estimation using randomized measurements~\cite{Ohliger1204.5735, Elben2203.11374, Notarnicola2112.11046}.


The rest of the paper is organized as follows.
In \secref{sec:SC_lattice}, we define the type-I and type-II strange correlators. With exact calculations on stabilizer Hamiltonians, and numerical computations for models away from the exactly soluble limit, we illustrate that under selective decoherence, the type-I strange correlator may be short-ranged, but the type-II strange correlator stays nontrivial and retain the memory of an underlying SPT phase. 
Furthermore, using the example of the 2d cluster state, we show that the type-I strange correlator can stay nontrivial for weak decoherence and undergo a transition into a short-ranged phase at a critical decoherence strength.

In \secref{sec:field_theory}, we evaluate both the type-I and type-II strange correlators using effective field theory descriptions for generic $1d$ and $2d$ SPT states. In this formalism, the strange correlators of decohered SPT states at the $d$ spatial dimension reduce to the ordinary correlation functions in the space-time of the boundary of the SPT state, and the two opposite boundaries are coupled by interactions in the selected channel. 

In \secref{sec:doubled}, we provide a general definition and formalism of symmetry-protected topological ensembles (SPT ensembles): we map a mixed state density matrix into a pure state in the doubled Hilbert space using the so-called Choi-Jamiołkowski isomorphism \cite{JAMIOLKOWSKI1972, CHOI1975}. Then a nontrivial SPT ensemble is mapped to a nontrivial pure quantum state in the doubled Hilbert space. We also argue that the type-II strange correlator can be viewed as a tool to decode the key information of a quantum state transmitted through a noisy channel. 

In \secref{sec:measure}, we establish the relation between the type-I strange correlators and more general non-local order parameters, such as the string-order parameter in the Haldane phase, showing that the type-I strange correlators provide an upper-bound for non-local order parameters. We also propose a method to experimentally probe the type-I strange correlator.

\section{Strange Correlators} \label{sec:SC_lattice}


In this section, we introduce the notion of strange correlators to identify SPT phases for mixed states. As we shall see later, under decoherence, a conventional diagnosis to measure SPT features, such as nonlocal order parameters, would decay exponentially with its support and thus cannot serve as a proper way to identify SPT physics. Therefore, a new diagnosis or even definition for symmetry-protected mixed state is required, and as we will show, strange correlators can serve as a new way to quantify the SPT features even in mixed states.
In order to illustrate the concepts, we would calculate strange correlators for several SPT states subject to decoherence. The physical interpretations of two different strange correlators will be discussed in \secref{sec:doubled} and \secref{sec:measure}.

Here we would like to stress that the strange correlators serve as a {\it diagnosis} for density matrices, but the actual definition for SPT ensemble is given in \secref{sec:doubled}, in the doubled Hilbert space formalism.

\subsection{Basic formalism}

Let us first elaborate on the basic formalism used in the paper.
Decoherence is defined as the process that incurs the loss of quantum coherence, evolving pure states into mixed states. 
Throughout this paper, we consider local decoherence models described by the quantum channel $\cE_i$ in the Kraus representation:
\begin{equation} \label{eq:noise_Kraus}
    \cE_i[\rho] = \sum_m  K^\vdagger_{m,i} \rho K_{m,i}^\dagger
\end{equation}
where $K_{m,i}$ is a Kraus operator with local support in the neighborhood of the $i$-th site, satisfying the trace-preserving condition $\sum_m K_{m,i}^\dagger K^\vdagger_{m,i} =1$. The global decoherence model would be defined by the composition of local decoherence models, $\cE = \prod_i \cE_i$.

Once we use the density matrix to describe a pure state that is invariant under certain symmetry $G$, the density matrix enjoys a ``doubled'' symmetry transformation, $i.e.$ the density matrix $\rho = |\Psi\rangle \langle \Psi |$ is invariant under separate ``left'' and ``right'' multiplication of the symmetry transformation, as $|\Psi\rangle$ is invariant under $G$: $\rho = g_L \rho g^\dagger_R$, $g_L, g_R \in G$. 
However, if we consider a density matrix of a mixed state or a pure state under a quantum channel in \eqnref{eq:noise_Kraus}, the notion of symmetry may be weakened to the density matrix being invariant under the adjoint action of the symmetry from both left and right $\rho=g \rho g^\dagger$ (for $g\in G$). 
In Ref.~\cite{Groot2021}, the former is defined as \emph{strong} symmetry condition and the latter as \emph{weak} symmetry condition. 
In fact, when it comes to the symmetry of density matrices, the weak symmetry condition is equivalent to the average symmetry defined in  Ref.~\cite{MaWang2022}. The average symmetry $G$ is defined for a random ensemble $\{\ket{\Psi}\}$ of states subject to a probability distribution $P(\ket{\Psi})$ that is invariant under the symmetry transformation, i.e. $P(g\ket{\Psi})=P(\ket{\Psi})$ for $g\in G$, even though $g\ket{\Psi}\neq\ket{\Psi}$. This directly implies that the density matrix $\rho=\sum_{\Psi} P(\ket{\Psi})\ket{\Psi}\bra{\Psi}$ is only invariant under the adjoint action $\rho=g \rho g^\dagger$ of the average symmetry as we introduced above. The disordered ensemble considered in Ref.~\cite{MaWang2022} also belongs to a set of general decohered density matrices. The density matrix for the disordered ensemble is obtained from a pure SPT state by applying a specific \emph{mixed unitary} quantum channel~\cite{mixedunitary} where Kraus operators satisfy certain constraints~\footnote{One requires that $(i)$ $K_m$ is unitary and $(ii)$ $K_m | \Psi_\spt\rangle$ should be connected to $|\Psi_\spt \rangle$ via an adiabatic path respecting exact symmetry for each Kraus operator $K_m$.}.  
However, a completely general decoherence channel does not have to satisfy such constraints.

In order to proceed further, we introduce the notion of weak and strong symmetries for quantum channels. First, consider a symmetry group $G = \{g \}$. We define a unitary channel ${\cal U}_g$ such that ${\cal U}_g[\rho] = U_g \rho U_g^\dagger$, where $U_g$ is the unitary representation of $g \in G$. Now, the decoherence channel $\cE$ described by a set of Kraus operators $\{K_i \}$ is \emph{weakly} symmetric in $G$ if
\begin{align}
    \cE \circ {\cal U}_g = {\cal U}_g \circ \cE
\end{align}
where $U_g$ is the unitary representation of $g$. On the other hand, we define a decoherence channel $\cE$ is \emph{strongly} symmetric in $G$ if
\begin{align}
    \cE[\rho U_g] = e^{i \theta_g} \cE[\rho] U_g, \quad  \cE[U_g \rho] = e^{-i \theta_g} U_g  \cE[\rho ]
\end{align}
for some $\theta_g$. This is equivalent to the condition for Kraus operators $U_g K_i U^\dagger_g = e^{i \theta_g} K_i$ for all $K_i$ and $g \in G$. These definitions are consistent with the aforementioned weak and strong symmetry conditions for a density matrix. These conditions can be also defined through the dynamics of the system under interaction with the environment as in Ref.~\cite{Groot2021}.

For some of the prototypical SPT states considered in this work, the system is defined with two symmetries, $G_A$ and $G_B$, and the ground state wave function of the system $|\Psi\rangle$ is symmetric under $G_A$ and $G_B$, meaning its density matrix is invariant under the separate left and right transformation of $G_A$ and $G_B$. 
Throughout the work, we often introduce a decoherence channel that is strongly symmetric in $G_A$ but weakly symmetric in $G_B$. This implies that the mixed density matrix under decoherence, denoted by $\rho^D = \cE[\rho]$ is invariant under the separate left and right transformations of $G_A$, but not necessarily for $G_B$.


\subsection{Review on Strange Correlator}

Let $\ket{\Psi}$ be a nontrivial SPT state, and $\ket{\Omega}$ be a trivial disordered state, then the following quantity
\begin{equation} \label{eq:strange_basic}
    C(r) \equiv \frac{\langle \Psi | O(0) O(r) | \Omega \rangle  }{\langle \Psi | \Omega \rangle }
\end{equation}
is called the \emph{strange correlator}, and it is expected to either saturate to a non-zero constant value or decay as power-law, for SPT states in $1d$ and $2d$~\cite{YouXu2013}. $O$ is an operator that transforms nontrivially under symmetries that define the SPT states. For noninteracting or weakly interaction fermionic topological insulators (TI) and topological superconductors (TSC), the strange correlators in higher dimensions should also decay with a power law. 

It is helpful to write the strange correlator in a slightly different form: $ C(r) \equiv \langle \Psi | \hat{C}(r) | \Psi \rangle$, where 
\beqn \hat{C}(r) = \frac{1}{|\langle \Psi | \Omega \rangle|^2} \Big[ O(0) O(r) |\Omega\rangle \langle \Omega| \Big]. 
\eeqn 
Hence the strange correlator can be viewed as the expectation value of an ``order parameter'' $\hat{C}(r)$ of $1d$ and $2d$ SPT phases, and the nontrivial behavior of this order parameter, either long-ranged or power-law decaying in its expectation value, can be viewed as a defining feature of an SPT wave function $|\Psi\rangle$. For a more comprehensive review for strange correlators of pure state SPT phases, please refer to Appendix D. 

\subsection{$1d$ Cluster State} \label{sec:1dcluster}

As an example, let us consider a $1d$ cluster state~\cite{1Dcluster_GHZ} with $2N$ sites, defined by the stabilizer Hamiltonian with $\mathbb{Z}_2  \times \mathbb{Z}_2 $ symmetry:
\begin{equation} \label{eq:1d_cluster_ham}
    H = - \sum_i Z_{i-1} X_i Z_{i+1},
\end{equation}
where the symmetry action is defined by the product of $X$ on even/odd sublattices. The periodic boundary condition is assumed, such that the Hamiltonian has a unique SPT ground state without degeneracies arising from boundary modes. We remark that under open boundaries, the system has anomalous boundary zero modes (spin-1/2) at each end, whose degeneracy is protected by symmetries; this is often considered to be a defining feature of the SPT state.

To evaluate a strange correlator, we use the following product state for a trivial disordered state: $\ket{\Omega} = \ket{+}^{\otimes 2N}$ for a $1d$ system with $2N$ sites, where $\ket{+}$ denotes the $X_n=+1$ eigenstate on every site.
For the strange correlator of $\mathbb{Z}_2^\todd$ charged operators separated by the $2n$ lattice spacings, we get
\begin{align} \label{eq:1d_cluster_even}
     C_\todd(2n) &= \frac{\langle \Psi | Z_1 Z_{2n+1} | \Omega \rangle  }{\langle \Psi | \Omega \rangle } \nonumber \\
     &= \frac{\langle \Psi | \prod_{m=1}^n X_{2m} | \Omega \rangle  }{\langle \Psi | \Omega \rangle } = 1,
\end{align}
where we have used $Z_1\prod_{m=1}^{n}X_{2m}Z_{2n+1}\ket{\Psi}=\ket{\Psi}$ for the SPT state $\ket{\Psi}$. This result is expected from the presence of a spin-1/2 zero mode at the boundary in the space-time rotated picture of the strange correlator. 

Similarly, the strange correlators of $\mathbb{Z}_2^\teven$ charged operators $Z_0$ and $Z_{2n}$ take a unit value, where we identify $Z_0 \equiv Z_{2N}$. On the other hand, if we replace $|\Psi\rangle$ with a trivial product state, strange correlators would vanish. 
Later in \secref{sec:measure} we will show that the strange correlator for the stabilizer Hamiltonian discussed here is directly connected to the well-known string order parameter of the Haldane phase~\cite{SOP1989, SOP2008}; therefore, the strange correlator can be indeed viewed as an ``order parameter'' of the SPT phase.

\subsection{Decoherence}

The strange correlators probe nontrivial information of the SPT state. What would happen to strange correlators if the SPT state is decohered through the noise channel that destroys, say the $\mathbb{Z}_2^\teven$ symmetry? First of all, we remark that as we discuss decoherence, we should use the density matrix formalism. In this case, the strange correlator expression in \eqnref{eq:strange_basic} should generalize as the following:
\begin{equation} \label{eq:strange_typeI}
    C^\textrm{I}(r) \equiv \frac{\tr(\rho_\spt O(0) O(r) \rho_0)}{\tr(\rho_\spt \rho_0)} 
\end{equation}
where $\rho_\spt$ and $\rho_0$ are density matrices of the decohered SPT and trivial states, respectively, and the superscript ``$\textrm{I}$'' stands for the ``type-I'' strange correlator. In the pure state limit, the above expression reduces into \eqnref{eq:strange_basic}. 

Now, consider a noise channel defined as the composition of local noises:
\begin{align} \label{eq:noise_basic}
    &\mathcal{E}_i: \rho \rightarrow (1-p) \rho  + p Z_i \rho Z_i, \,\, \cE \equiv \cE_2 \circ \cE_4 \circ \cdots \cE_{2N} \nonumber \\
    &\Rightarrow \cE[\rho] = \sum_{\bm{\eta}}P(\bm{\eta})\,  K^\dagger_{\bm{\eta}} \rho K^\vdagger_{\bm{\eta}},\quad K_{\bm{\eta}} \equiv \prod_m Z_{2m}^{\eta_{2m}}.
\end{align}
where $p$ is the probability of having a dephasing noise locally, $\bmeta \equiv (\eta_2, \eta_4, ..., \eta_{2N})$ is a bit-string of $\{0,1\}$ characterizing the $Z$ noise operator $K_{\bm{\eta}}$, and  $P(\bm{\eta})=\prod_{m=1}^{N} P(\eta_{2m})$ with $P(0)=1-p$ and $P(1)=p$ is the probability of having the noise $\bm{\eta}$. Although this noise channel perturbs the system to locally break the $\mathbb{Z}_2^\teven$ symmetry along every quantum trajectory, the channel is weakly symmetric in $\mathbb{Z}_2^\textrm{even}$. Since the channel does not act on odd sites, it is strongly symmetric in $\mathbb{Z}_2^\textrm{odd}$.

Under this channel, the pure SPT state becomes a mixed state ensemble, denoted by $\rho^D_\spt \equiv \cE[\rho_\spt]$. 
Now we evaluate the type-I strange correlator of  $\rho^D_\spt$ against the trivial state density matrix $\rho_0 = |\Omega \rangle \langle \Omega |$. We first evaluate the denominator of \eqnref{eq:strange_typeI}:
\begin{align}
    \textrm{tr}\big(\rho^D_\spt \rho^\vdagger_0\big) &= \sum_{ \bm{\eta} } P(\bm{\eta})  |{\langle \Psi| K_{\bm{\eta}} | \Omega \rangle }|^2 = \frac{1}{2} | {\langle \Psi | \Omega \rangle } |^2.
\end{align}
This is because odd number of $Z$ operators can change the $\mathbb{Z}_2^\teven$ parity, whose expectation value vanishes between two states with the same parity. Such a case corresponds to $\sum_m \eta_{2m} \equiv 1 \mod 2$, which happens about half the time in the limit $N \rightarrow \infty$. Interestingly, this implies that we can use the trivial product state $\ket{\Omega}$ with an arbitrary parity, since the decoherence would flip the parity of the SPT state half the time, i.e., $\cE[\rho_\spt] = \frac{1}{2} \rho_\spt^\textrm{e} + \frac{1}{2}\rho_\spt^\textrm{o}$ in the limit $N \rightarrow \infty$ where $\tr \rho_\spt^\textrm{e,o} =1$. For an even parity error $K_{\bm{\eta}}$ such as $Z_{2a} Z_{2b}$, it satisfies $K_\bmeta \ket{\Psi} = \prod_{m=a}^{b-1} X_{2m+1} \ket{\Psi}$
and then the action of the product of $X$ on $\langle \Omega|$ should square to one. 

Similarly, the numerator of the strange correlator for the $\mathbb{Z}_2^\todd$ charged operator is
\begin{align}
    &\textrm{tr}\big(\rho^D_\spt Z^\vdagger_1 Z^\vdagger_{2n+1} \rho^\vdagger_0\big) \nonumber \\
    & \quad = \sum_{ \bm{\eta} } P(\bm{\eta}) {\langle \Psi| K_{\bm{\eta}} Z_1 Z_{2n+1} | \Omega \rangle }  {\langle \Omega| K^\dagger_{\bm{\eta}} | \Psi \rangle } \nonumber \\
    & \quad = \sum_{ \bm{\eta} } P(\bm{\eta}) {\langle \Psi|  \prod_{m=1}^n X_{2m} K_\bmeta | \Omega \rangle }  {\langle \Omega| K^\dagger_{\bm{\eta}} | \Psi \rangle } \nonumber \\
    &\quad = \frac{1}{2} | {\langle \Psi | K_\bmeta | \Omega \rangle } |^2  \Big[ \prod_{m=1}^n  \sum_{ \eta_{2m} }    (-1)^{\eta_{2m}} P(\eta_{2m}) \,  \Big]  \nonumber \\
    & \quad = \frac{1}{2} | {\langle \Psi | \Omega \rangle } |^2 e^{-n/\xi},\,\,\,\,\,\, \xi = 1/\ln(1/(1-2p)).
\end{align}
However, for the strange correlator of $\mathbb{Z}_2^\teven$ charged operators, its numerator is
\begin{equation}
    \tr(\rho^D_\spt Z_0 Z_{2n} \rho_0)=\tr(\rho^D_\spt  \rho_0)=\frac{1}{2}| {\langle \Psi | \Omega \rangle } |^2,
\end{equation} because the operator $Z_0 Z_{2n}$ can commute through $K_{\bm{\eta}}$ to act on the state $\bra{\Psi}$ and become $\prod_{m=1}^{n}X_{2m-1}$ to commute back and hit $\ket{\Omega}$.
Following \eqnref{eq:strange_typeI}, we obtain 
\begin{equation}\label{eq:strange_typeI_result}
    C^\textrm{I}_\teven(2n) = 1, \qquad C^\textrm{I}_\todd(2n) =  e^{-n/\xi}.
\end{equation}
for the SPT state decohered under $\mathbb{Z}_2^\teven$ noise channel. Therefore, the odd-sited strange correlator becomes exponentially decaying, and its length scale depends on the strength of the decoherence. For $p \ll 1$, we see that $\xi$ is very large, and the strange correlator would look the same as a pure state SPT phase for small system size. However, in the thermodynamic limit, the odd-sited strange correlator is always exponentially decaying to zero while the even-sited strange correlator behaves the same as the pure state.  

A recent work pointed out that the notion of SPT order can still be defined even if the symmetry only holds in the average sense in both disordered ensemble and quantum channels~\cite{MaWang2022}. Under decoherence, although the $\mathbb{Z}_2^\teven$ symmetry is broken for every quantum trajectory, the ensemble of all trajectories still preserves the symmetry on average, thus an ``average SPT'' order (or mixed-state SPT order) should still be expected. 
Our result \eqnref{eq:strange_typeI_result} suggests that although some type-I correlators (such as $C^\textrm{I}_\teven$) can exhibit non-trivial behavior in the mixed-state SPT phase, in general, they are insufficient to detect the full nature of the SPT order.

{In the pure state limit, the SPT order is manifested by the \textit{mutual} charge-flux attachment for both the $\mathbb{Z}_2^\teven$ and $\mathbb{Z}_2^\todd$ symmetries. However, in the mixed-state case, the type-I strange correlator only remains nontrivial for $C^\textrm{I}_\teven$ associated with the $\mathbb{Z}_2^\teven$ symmetry charge, but not for $C^\textrm{I}_\todd$ associated with the $\mathbb{Z}_2^\todd$ symmetry charge. To find a more comprehensive diagnosis for the mutual SPT response,} one can define a new type of strange correlator as follows:
\begin{equation} \label{eq:strange_typeII}
    C^\textrm{II}(r) \equiv   \frac{\tr(\rho^D_\spt O(0) O(r) \rho^\vdagger_0 O(r)^\dagger O(0)^\dagger )}{\tr(\rho^D_\spt \rho^\vdagger_0)}
\end{equation}
which is called the ``type-II'' strange correlator. Then, following the same calculation as above, one can show that for $O=Z$,
\begin{equation}
    C^\textrm{II}_\teven(2n) = C^\textrm{II}_\todd(2n) = 1.
\end{equation}
which is nontrivial for both even and odd sites. This is because the noise only flips the sign of the operator overlap between $|\Omega\rangle$ and $K_\bmeta |\Psi \rangle$, which squares to one in the type-II strange correlator evaluation. Hence the type-II strange correlator is analogous to the Edwards-Anderson correlator of the type-I strange correlator. For a $\mathbb{Z}_n$ generalization of this framework, see \appref{app:example}.

\begin{figure}[!t]
    \centering
    \includegraphics[width = 0.49 \textwidth]{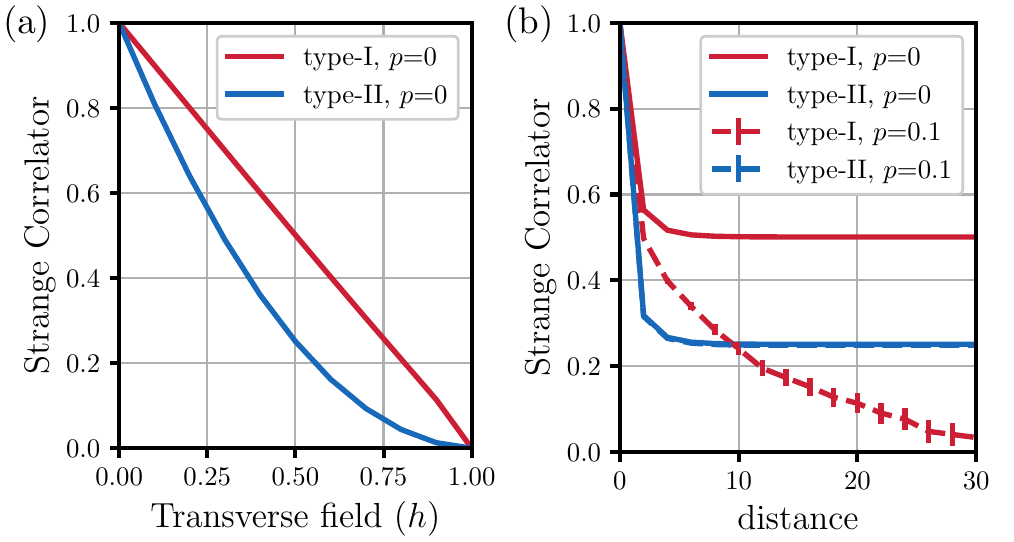}
    \caption{\label{fig:1d_numerics} {\bf Strange Correlators in 1d Cluster State.} The DMRG numerics is performed at the system size $L=100$ for $H = -\sum_n (Z_{n-1} X_n Z_{n+1} + h X_n)$ under periodic boundary condition.
    (a) Odd-sited strange correlators at the distance $L/2$ as functions of the transverse field $h$. (b) Odd-sited strange correlators as functions of the distance at $h=1/2$. Red (blue) curves represent type-I (II) strange correlators. Solid (dashed) lines are obtained at the decoherence strength $p=0$ ($p=0.1$).  Here, the decohered density matrix is obtained for the ensemble of $4 \times 10^3$ different disorder realizations on even sites. We remark that the type-II strange correlator at $p=0.1$ stays exactly on top of the value at $p=0$ with vanishingly small sample variance.
    } 
\end{figure}

\subsection{Away from the fixed point}

So far, we have discussed the strange correlators for the $1d$ $\mathbb{Z}_2 \times \mathbb{Z}_2$ SPT state in the stabilizer limit with a zero correlation length ($\xi=0$). Even away from the stabilizer limit, the aforementioned properties of the type-I and type-II strange correlators generally hold: while the type-I strange correlator would decay exponentially with distance under decoherence, the type-II strange correlator would remain long-ranged under decoherence. This behavior will be more systematically discussed in the later section using field theory.

To demonstrate the behavior, consider introducing a transverse field $-h \sum_n X_n$ to the Hamiltonian in \eqnref{eq:1d_cluster_ham}, which drives the ground state away from the soluble limit. 
In \figref{fig:1d_numerics}, we numerically obtained both types of $\mathbb{Z}_2^\todd$-charged strange correlators $C^\textrm{I}_\todd$ and $C^\textrm{II}_\todd$ computed against $|+\rangle^{\otimes L}$ as functions of $r$ at $h=0.5$, with and without decoherence on the even sites as in \eqnref{eq:noise_basic}. 
The plot illustrates two important points: $(i)$ away from the fixed point, the magnitudes of the strange correlators decrease, which completely vanish at the known topological-trivial transition point $h=1$; and $(ii)$ under decoherence, while $C^\textrm{I}$ decays exponentially with distance, $C^\textrm{II}$ stays robust. Therefore, the type-II strange correlator serves as a probe for the SPT physics under decoherence. As a side remark, we found that the magnitude of the type-I strange correlator in each quantum trajectory is very close to each other; the main difference among the type-I strange correlators in different quantum trajectories is their signs. 

Although not shown in the figure, we remark that for $h>0$, both $C^\textrm{I}_\teven$ and $C^\textrm{II}_\teven$ may increase under the decoherence on even sites. This is because while $C^\textrm{I,II}_\teven$ probe correlations between exactly localized charged operators, away from the fixed point, $\mathbb{Z}_2$-charged operator become diffused across the correlation length $\xi$. As decoherence effectively diffuses a localized charged operator $Z$ over multiple sites, decoherence can help enhance the strange correlator. 

\subsection{Generic Noise Model}

One can consider more generic decoherence channels as defined in \eqnref{eq:noise_Kraus}. For example, one may consider the following depolarization channel on site $i$
\begin{equation} \label{eq:generic_noise}
    \mathcal{E}_i: \rho \rightarrow (1-p) \rho  + \frac{p}{3} (X_i \rho X_i + Y_i \rho Y_i + Z_i \rho Z_i).
\end{equation}
When we turn on this noise on all the even sites, it is straightforward to see that the behavior of the type-I and type-II strange correlators would remain the same as was discussed in the previous sections.  

Instead, one may also consider a noise model coherent over multiple sites, i.e., Kraus operators $K_i$s in \eqnref{eq:noise_Kraus} are extended over multiple sites. In this case, the global decoherence channel acts as a stochastic random short-depth local unitary transformation. While doing so, we can still require the noise channel to respect the doubled symmetry of $\mathbb{Z}_2^\todd$, and the adjoint action of $\mathbb{Z}_2^\teven$. For each quantum trajectory, such a noise action can be decomposed into strictly a local noise in \eqnref{eq:noise_basic} and short-depth local unitary transformations. Since a unitary transformation can decrease the magnitude of the type-II strange correlator as shown in \figref{fig:1d_numerics}(b), a generic noise model can change the magnitude of the type-II strange correlator.

Finally, we remark two additional cases: (i) If $\cE$ is strongly symmetric under $G = \mathbb{Z}_2 \times \mathbb{Z}_2$ (does not break any symmetry of an underlying SPT state), an observable (e.g. strange correlators or non-local order parameters) that probes the presence of an SPT order remains unchanged. 
(ii) If $\cE$ is weakly symmetric under $G = \mathbb{Z}_2 \times \mathbb{Z}_2$ (break the entire symmetry of an underlying SPT state), both types of strange correlators would become immediately trivial in this case. However, as we later discuss, if protecting symmetries are of higher forms, emergent symmetries can appear to still protect strange correlators.

\subsection{2d Cluster State}

A 2d generalization of the cluster state~\cite{2Dcluster, 2Dcluster_toric} in the Lieb lattice is an SPT state~\cite{Yoshida2016} exhibiting the mixed topological response (often ``anomaly'' in literature) between 0-form $\mathbb{Z}_2^{(0)}$ and 1-form $\mathbb{Z}_2^{(1)}$ symmetries defined on the square lattice, where qubits reside on both vertices and edges. For a $L \times L$ square lattice, there are $N_v = L^2$ vertex qubits and $N_e = 2L^2$ edge qubits. Its stabilizer Hamiltonian is defined as
\begin{equation} \label{eq:2dclusterHam}
    H = - \sum_{v} \qty( X_v \prod_{e \in \mathrm{d}v } \bm{Z}_e ) - \sum_{e} \qty( \bm{X}_e \prod_{v \in \partial e} Z_v ),
\end{equation}
where $\partial$ stands for the boundary operator on the lattice and $\mathrm{d}=\star\partial\star$ is the coboundary operator (with $\star$ being the Hodge dual). Bold symbols $\bm{Z}$ and $\bm{X}$ act on edges, and unbold symbols $Z$ and $X$ act on vertices. Here, all terms in the Hamiltonian commute with one another, and the groundstate satisfies that each term be 1. This implies that $B_p \equiv \prod_{e \in \partial p} \bm{X}_e=1$ for any plaquette $p$. We denote the symmetry group by $G_A \equiv \mathbb{Z}_2^{(0)}$ and $G_B \equiv \mathbb{Z}_2^{(1)}$, where the 0-form symmetry charge (generator) $g \equiv \prod_{v} X_v \in G_A$ and the 1-form symmetry charge $h_\gamma \equiv \prod_{e \in \gamma} \bm{X}_e \in G_B$ for any closed loop $\gamma$ along the bonds. Again, the ground state of \eqnref{eq:2dclusterHam} has the decorated domain wall (defect) structure: the defect of the 1-form symmetry measured by $\prod_{e \ni v} \bm{Z}_e$ is bound to the charge of the 0-form symmetry measured by $X_v$; also the defect (domain wall) of the 0-form symmetry measured by $Z_v Z_{v'}$ is bound to the charge of the 1-form symmetry measured by $X_e$. As shown in \cite{Yoshida2016}, the mixed topological response between $\mathbb{Z}_2$ 0-form and $\mathbb{Z}_2$ 1-form symmetries in the SPT state enforces the boundary to be nontrivial under open boundary condition; when vertex qubits are exposed, the boundary should spontaneously break the 0-form symmetry, forming a $\mathbb{Z}_2$ ferromagnet. 

Now, we are ready to calculate the strange correlators for this state. There are two operators we can inspect: $Z_v Z_{v'}$ which is the correlation of $G_A$-charged operators, and $\prod_{e \in \gamma^\star} \bm{Z}_e$ which is the Wilson loop of $G_B$-charged operators where $\gamma^\star$ is a closed loop on the dual lattice. 

First, we evaluate strange correlators for the stabilizer state. Let $\rho_0 = | \Omega \rangle \langle \Omega |$, where $|\Omega \rangle = |+\rangle^{\otimes(N_e+N_v)}$. Then,
\begin{align}
    C_{A}^\textrm{I} \equiv  \frac{\tr(\rho_\spt Z_v Z_{v'} \rho_0)}{\tr(\rho_\spt \rho_0)} &= \frac{\tr(\rho_\spt \prod_{e \in l} \bm{X}_e \rho_0)}{\tr(\rho_\spt \rho_0)} = 1
\end{align}
where $l$ is an open string connecting two vertices $v$ and $v'$. Since $B_p = 1$ for both $\ket{\Psi}$ and $\ket{\Omega}$, the RHS is independent of the choice of $l$. Similarly, for any closed loop $\gamma = \rd \cA$, where $\cA$ is the region enclosed by the loop $\gamma$,
\begin{align}
    C_{B}^\textrm{I} &\equiv  \frac{\textrm{tr}(\rho_\spt \prod_{e \in \rd \cA} \bm{Z}_e \rho_0)}{\tr(\rho_\spt \rho_0)} \nonumber \\
    &= \frac{\tr(\rho_\spt \prod_{v \in \cA} X_v \rho_0)}{\tr(\rho_\spt \rho_0)} = 1
\end{align}
Accordingly, the type-II strange correlators would simply be $C_A^\textrm{II} = C_B^\textrm{II} = 1$.

Interestingly, under decoherence, $C_A$ and $C_B$ exhibit a qualitative difference. For example, consider the noise channel $\cE_B$ of the form in \eqnref{eq:noise_basic} acting on edge qubits, which is weakly symmetric in $G_B = \mathbb{Z}_2^\textrm{(1)}$.
To facilitate the calculation of strange correlators, it is convenient to define the projection operator $\cP$ for vertex and edge qubits as the following:
\begin{align} \label{eq:projector}
    \cP_0^v &= \prod_{v} \frac{1 + X_v}{2}, \qquad \cP_0^e = \prod_{e} \frac{1 + \bm{X}_e}{2}.
\end{align}
Then, the trivial state density matrix is given by the product of two projectors, $\rho_0 = \cP_0^v \otimes \cP_0^e$. Using this,
\begin{align}
    \tr( \cE_B[\rho_\spt] \rho_0) &= \tr( \rho_\spt \cE_B[\rho_0]) = \langle \Psi | \cP^v_0 \otimes \cP^e_\cE | \Psi \rangle, \nonumber \\
    \cP^e_\cE  \equiv \cE[\cP^e_0] &= \prod_{e} \frac{1}{2} (1 + (1-p) \bm{X}_e +  p \bm{Z}_e \bm{X}_e \bm{Z}_e) 
\end{align}
where we use the self-adjointness of the decoherence channel in the first line. Note that $\cP_\cE^e = \prod_{e} \frac{1}{2} (1 + (1-2p) \bm{X}_e )$ since $ZXZ = -X$. By expanding the above expressions and using the Lemma in \appref{app:note}, we obtain that
\begin{align} \label{eq:1form_partition}
    \textrm{tr}\big( \rho_\spt^D \rho^\vdagger_0 \big) &= \frac{2 }{2^{N_e} \cdot 2^{N_v} } \sum_{\gamma} (1-2p)^{|\gamma|}
\end{align}
where the summation over $\gamma$
is taken for all closed loops. This is exactly the high-temperature expansion of the Ising model partition function, ${\cal Z}:=\sum_{\{ \sigma \}} e^{-\beta \sum_{\langle ij \rangle} \sigma_i \sigma_j }$. Similarly, 
\begin{align} \label{eq:1form_corr}
    &\tr( \cE_B[\rho_\spt] Z_v Z_{v'} \rho_0) = \tr( \rho_\spt Z_v Z_{v'}  \cE_B[\rho_0]) \nonumber \\
    & = \langle \Psi | \prod_{e \in l} \bm{X}_e \cP^v_0 \otimes \cP^e_\cE | \Psi \rangle\nonumber\\
    &= \frac{2 }{2^{N_e}  2^{N_v} } \sum_{\gamma'} (1-2p)^{|\gamma'|}
\end{align}
where the summation over $\gamma'$ is taken over all possible loops added by the open string $l$ (they are $\mathbb{Z}_2$-valued). The channel $\cE_B$ could be moved to $\rho_0$ since it acts trivially on the vertices. 
This is equivalent to the high-temperature expansion of the Ising correlation function. Therefore, we get
\begin{align} \label{eq:1form_strangeIA}
    C_A^\textrm{I} = \frac{\textrm{tr}( \rho^D_\spt Z_v Z_{v'} \rho_0)}{ \textrm{tr}( \rho^D_\spt \rho_0) } = \expval{Z_v Z_{v'}}_\beta^\textrm{2dIsing},
\end{align}
the correlation function of classical 2d ferromagnetic Ising model at the inverse temperature $\beta = \tanh^{-1}(1-2p)$. We emphasize that for a different choice of the reference state $|\Omega \rangle$, $C_A^\textrm{I}$ will be the correlation function of the random bond Ising model, see \eqnref{eq:RBIM_Ising}. 

To compute the type-II strange correlator, we calculate its numerator as the following:
\begin{align} \label{eq:1form_corr_SCII}
    &\tr( \cE_B[\rho_\spt] Z_v Z_{v'} \rho_0 Z_v Z_{v'}) \nonumber\\
    &= \tr( \rho_\spt Z_v Z_{v'}  \cE_B[\rho_0] Z_v Z_{v'}) \nonumber \\
    & = \langle \Psi | \prod_{e \in l} \bm{X}_e \cP^v_0 \otimes \cP^e_\cE \prod_{e \in l} \bm{X}_e | \Psi \rangle \nonumber\\
    &= \textrm{tr}\big( \rho_\spt^D \rho_0 \big).
\end{align}
Therefore, $C^\textrm{II}_A = 1$. Note that $C_B^\textrm{I,II}$ are unaffected by the noise $\cE_B$, taking a unit value.  

It is instructive to understand how the conventional non-local order parameter for the 2d SPT state behaves. In the 2d SPT pure state, the following string order parameter takes a finite value in the limit where its length diverges~\cite{Yoshida2016}:
\begin{equation}
    \lim_{|l| \rightarrow \infty} \Big\langle Z_v \Big[\prod_{e \in l} \bm{X}_e \Big] Z_{v'} \Big \rangle \neq 0,
\end{equation}
where $l$ is an open string along the bonds and $v,v'$ are two vertices where the string ends. This can be easily understood if we imagine applying a symmetric quantum circuit $U$ that moves the state away from the fixed point wave function. Since $U$ commutes with a string of $\bm{X}$, it only acts nontrivially on $Z_{v,v'}$. Accordingly, for any order parameter of the above form, it will be corrected by a constant amount as long as one is within the same phase.
However, this order parameter is actually short-ranged for the decohered mixed state. Using the above formalism, it is straightforward to show that
\begin{align} \label{eq:2dSPT_NLO_decoherence}
    &\textrm{tr}\Big( \cE_B[\rho_\spt] Z_v \Big[\prod_{e \in l} \bm{X}_e \Big] Z_{v'} \Big) \nonumber \\
    &\quad = (1-2p)^{\abs{l}} \textrm{tr}\Big( \rho_\spt Z_v \Big[\prod_{e \in l} \bm{X}_e \Big] Z_{v'} \Big)\nonumber\\
    &\quad= e^{-l/\xi}
\end{align}
where $\xi = 1/\ln(1/(1-2p))$. Therefore, it implies that the conventional non-local order parameter would fail to detect the nontrivial structure of the underlying SPT state if there is decoherence, while the type-I or type-II strange correlator can still detect in this case. This issue will be discussed more in the \secref{sec:measure}.

On the other hand, if we consider the noise channel $\cE_A$ of the form in \eqnref{eq:noise_basic} acting on vertex qubits, which is weakly symmetric in $G_A = \mathbb{Z}_2^\textrm{(0)}$, we can evaluate that 
\begin{align} \label{eq:1form_strangeIB}
     C_B^{\textrm{I}}(\rd \cA)  &= (1-2p)^{|\cA|}
\end{align}
This indicates that the type-I strange correlator probing the 1-form symmetry decays in an area-law manner, i.e., short-ranged. This result is consistent with the earlier discussion about $1d$ $\mathbb{Z}_2 \times \mathbb{Z}_2$ SPT state, as upon compactification into a cylindrical geometry which is effectively 1d, the 1-form symmetry becomes 0-form, and we have shown that the corresponding type-I strange correlator is short-ranged under the decoherence of the other symmetry.

\begin{figure}[!t]
    \centering
    \includegraphics[width = 0.45 \textwidth]{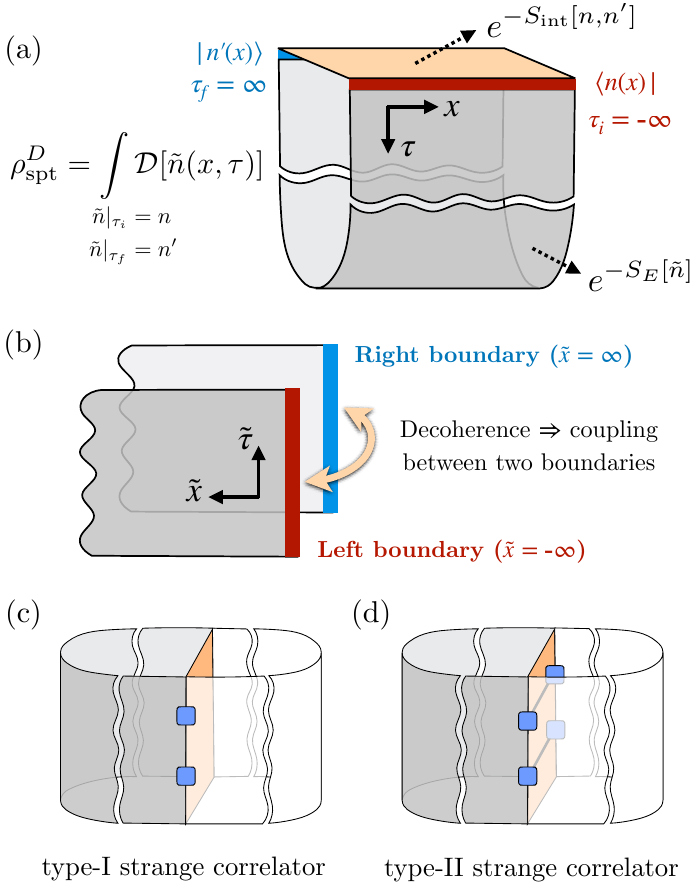}
    \caption{\label{fig:sc_pathintegral} {\bf Path Integral and Strange Correlators.} Diagrams for (a) The density matrix of a SPT state under decoherence as an imaginary-time path integral. The two boundaries at the temporal direction are decoupled for a pure state. However, decoherence acts like a coupling between two boundaries, collapsing the doubled symmetries down to its diagonal subgroup. (b) Under space-time rotation $(x,\tau) \rightarrow (\tilde{x},\tilde{\tau})$, the two boundaries along the temporal direction $\tau$ would become two opposite boundaries along the spatial direction $\tilde{x}$; also, decoherence acts as a perturbation connecting two spatial boundaries.  (c) The type-I strange correlator where blue squares corresponding to charged operators (order parameters) are present only at the $\tau = -\infty$ boundary. The grey (white) sheet corresponds to the path integral formulation of the SPT (trivial state). (d) The type-II strange correlator with charged operators is placed at both $\tau = \pm \infty$ boundaries. }
\end{figure}

\section{Effective Field Theory Evaluation} \label{sec:field_theory}

Many SPT phases classified and constructed through the group cohomology formalism in Ref.~\cite{wenspt,wenspt2} can be described by an effective field theory called nonlinear sigma model (NLSM)~\cite{ashvinsenthil2012,binlsm}. Various different SPT phases share similar physics captured by an NLSM with a topological $\Theta$-term. For example, $1d$ bosonic SPT phases can often be viewed as the descendants of the Haldane phase of a spin-1 chain, through reducing the SO(3) spin symmetry down to its subgroups, as long as the subgroup still has a nontrivial projective representation carried by the edge state. Hence all these $1d$ bosonic SPT states can be described as an O(3) NLSM in the $(1+1)d$ space-time with a $\Theta-$term at $\Theta = 2\pi$~\cite{haldane1}. At the physical boundary, the $\Theta-$term of the NLSM reduces into a WZW term in the $(0+1)d$ space-time~\cite{ng1994}, which gives us two-fold degenerate boundary states carrying a projective representation of the underlying symmetry group.  

The effective field theory description also makes the physical interpretation of the construction of these SPT states (e.g. decorated domain wall) transparent. Hence an effective field theory evaluation of the strange correlator of the SPT states under decoherence would be more universal, and is applicable to situations with continuous symmetries, as well as higher dimensions. For a self-contained review on field-theoretic descriptions of strange correlators and SPT phases, see \appref{app:NLSMreview}.

\subsection{$1d$ SPT states}\label{sec:1d field theory}

As was observed in Ref.~\cite{sptwf,YouXu2013}, the wave function of many bosonic SPT states in $1d$ can be inferred from the bulk topological $\Theta-$term of the NLSM, and written as 
\begin{align}
&\quad |\Psi \rangle \sim \int D[\vect{n}(x)] \exp\left( - \cS[\vect{n}(x)] \right) |\vect{n}(x) \rangle \nonumber \\ &\quad \cS = \int dx \ \frac{1}{g} (\nabla_x \vect{n}(x))^2 + \mathrm{WZW}[\vect{n}(x)] \nonumber \\ &\mathrm{WZW}[\vect{n}(x)] = \int_0^1 du \int dx \ \frac{2\pi \ii}{4\pi} \epsilon_{abc} \tilde{n}^a \partial_u \tilde{n}^b \partial_x \tilde{n}^c \label{1dwf}    
\end{align}
where $\vect{n}$ is a unit three component vector with length $|\vect{n}|\,{=}\,1$, and $\tilde{\vect{n}}(x, u)$is an extension of $\vect{n}(x)$ into the space $(x, u)$ where $\tilde{\vect{n}}(x, 0) = \vect{n}(x)$, $\tilde{\vect{n}}(x, 1) = \hat{z}$. This wave function can describe SPT phases in $1d$ with symmetries $\SO(3)$, $\U(1)\,{\rtimes}\,\mathbb{Z}_2$, $\mathbb{Z}_2\,{\times}\,\mathbb{Z}_2$, $\U(1)\,{\times}\,\mathbb{Z}_2^T$, $\mathbb{Z}_2^T$, etc. The Wess-Zumino-Witten term in \eqnref{1dwf} can be viewed as the termination of the bulk topological $\Theta-$term at a temporal boundary~\cite{sptwf}, which is the ``space-time dual'' of the WZW term at the physical real space boundary~\cite{ng1994} of the SPT state. Both the temporal boundary and spatial boundary would inherit 't Hooft anomaly from the bulk topology, and the strange correlator is one way to capture this boundary 't Hooft anomaly.

To make a connection with the computation based on the stabilizer Hamiltonian of the SPT states presented in the previous section, let us assume the symmetry of the system is $\mathbb{Z}_2^A \times \mathbb{Z}_2^B$, which acts on the vector $\vect{n}$ as \beqn && \mathbb{Z}_2^A: (n_x, n_y, n_z) \ra (- n_x, -n_y, n_z); \cr\cr && \mathbb{Z}_2^B: (n_x, n_y, n_z) \ra (n_x, -n_y, - n_z). \eeqn

The density matrix of the pure SPT state $|\Psi\rangle \langle \Psi|$ is given as the following in the basis of $|\vect{n}(x) \rangle$: 
\begin{align}
    \label{puredm}
&\rho_{\spt} \sim \int {\cal D}\{\vect{n}(x), \vect{n}'(x)\} \ e^{- \cS[\vect{n}] - \cS^\ast [\vect{n}'] } | \vect{n}(x) \rangle \langle \vect{n}'(x) |.
\end{align}
As a pure state density matrix, all the symmetries of the system have manifestly ``doubled'' as we explained in the introduction: this pure state density matrix is invariant under separate ``left'' and ``right'' symmetry transformations, which act on $\vect{n}(x)$ and $\vect{n}'(x)$ respectively. We would like to consider a mixed-state density matrix built based upon the SPT wave function $|\Psi\rangle$. One way to achieve this is increasing the ``weight'' of the diagonal elements of the density matrix to decrease the purity (or increase the number of non-zero Schmidt eigenvalues). This is equivalent to turning on some ``interaction'' $\cS^{\textrm{int}}[\vect{n}(x), \vect{n}'(x)]$ between $\vect{n}(x)$ and $\vect{n}'(x)$ in the density matrix, as diagrammatically represented in \figref{fig:sc_pathintegral}(a): 
\begin{align} \label{mixeddm}
    \rho^{D}_{\spt} &\sim \int \cD\{\vect{n}, \vect{n}' \}  e^{ - \cS[\vect{n}] - \cS^\ast [\vect{n}'] - \cS^{\textrm{int}}[\vect{n}, \vect{n}']  } | \vect{n}(x) \rangle \langle \vect{n}'(x) |. 
\end{align}
Since $\cS^{\textrm{int}}[\vect{n}(x), \vect{n}'(x)]$ should favor specific combinations of $\vect{n}(x)$ and $\vect{n}'(x))$, the system would be no longer invariant under all left and right symmetry transformations; still, it must remain invariant under the simultaneous left and right transformations on $\vect{n}(x)$ and $\vect{n}'(x)$ as long as $\cS^{\textrm{int}}$ is invariant under the simultaneous actions.

To proceed, we define the trivially disordered state $|\Omega\rangle$ as an equal weight superposition of all the configurations of $\vect{n}(x)$: \beqn |\Omega \rangle \sim \int D \vect{n}(x) \, |\vect{n}(x)\rangle. \eeqn Then, the type-I \eqref{eq:strange_typeI} and type-II \eqref{eq:strange_typeII} strange correlators are expressed in terms of vectors $\vect{n}$ as follows:
\begin{align}
    C^{\mathrm{I}}_{ab}(x) &=\frac{\mathrm{tr}\left( n_a (x) n_b(0)\rho^{D}_{\spt} \rho_0 \right)}{\mathrm{tr}\left(\rho^{D}_{\spt} \rho_0 \right)}.  \nonumber \\
    C^{\mathrm{II}}_{ab}(x) &=\frac{\mathrm{tr}\left( n_a(x) n_b(0)\rho^{D}_{\mathrm{spt}} n_a(x) n_b(0) \rho_0 \right)}{\mathrm{tr}\left(\rho^{D}_{\spt} \rho_0 \right)}. 
\end{align}

Formally, after the space-time rotation ($x \rightarrow \tau$) illustrated in \figref{fig:sc_pathintegral}(b), the original type-I strange correlator is mapped to the spin-spin correlation along the temporal direction of two interacting spin-1/2 degrees of freedom (the $0d$ boundary of a $1d$ SPT state), one from $\bm{n}$ and the other from $\bm{n}'$: 
\begin{equation} \label{scItemporal}
    C^\mathrm{I}_{ab}(x) \sim \langle n_a(x) n_b (0) \rangle \sim \langle S_a(\tau) S_b(0) \rangle,
\end{equation}
as in the path integral formalism, an isolated spin-1/2 is represented by a $(0+1)d$ NLSM with a WZW term at level-1. The evaluation of the type-II strange correlator will be mapped formally to an evaluation of the ``doubled'' temporal spin-spin correlation of the two interacting spins, $\vect{S}$ and $\vect{S}^\prime$: \beqn C^\mathrm{II}_{ab}(x) \sim \langle S_a(\tau) S_b(0) \ S^{\prime}_a(\tau) S^{\prime}_b(0) \rangle. \label{scIItemporal} \eeqn 
Here we would like to clarify that the mapping to the temporal spin correlation is a mathematical method of computing the strange correlator, which is a quantity associated with the bulk density matrix. 

As an example, let us introduce the following ``interaction'' in the density matrix
\begin{align}
    \mathcal{S}^{\textrm{int}}_{1d, 1}[\vect{n}(x), \vect{n}'(x)] \sim \int dx \ u \left( \vect{n}(x)\cdot \vect{n}'(x) \right). \label{int1d1}
\end{align}
Physically this interaction corresponds to introducing decoherence of all degrees of freedom of the system, as now the density matrix in \eqnref{mixeddm} is no longer invariant under any separate left or right $\mathbb{Z}_2^A$ or $\mathbb{Z}_2^B$ transformation, though it is still invariant under simultaneous left and right transformations; in other words, the decoherence introduced by $S^{\textrm{int}}_{1d,1}$ should be analogous to introducing temperature to the density matrix, which thermalizes all degrees of freedom. A similar idea of generating mixed density matrices has been explored for states constructed with loop degrees of freedom~\cite{chamon}. Most naturally, when the density matrix is driven into a mixed state under decoherence, the interaction $\mathcal{S}^{\textrm{int}}_{1d, 1}$ should favor the ``diagonal'' configurations $\vect{n}(x) \sim \vect{n}'(x)$, $i.e.$ we need $u < 0$ in \eqnref{int1d1}. When $\vect{n}(x) \sim \vect{n}'(x)$, the two WZW terms, $i.e.$ $\mathrm{WZW}[\vect{n}(x)]$ and $\mathrm{WZW}[\vect{n}'(x)]$ tend to cancel each other. Without the WZW term, both the type-I and type-II strange correlators would be short-range, for all components $a$, $b$. This is consistent with the picture under space-time rotation, as the interaction term $S^{\textrm{int}}_{1d,1}$ would translate into the following spin-spin interaction of the zero-dimensional system \beqn H^{\textrm{int}}_{0d,1} \sim J \vect{S} \cdot \vect{S}' + \cdots. \eeqn The ellipsis includes terms that reduce the SO(3) spin symmetry down to $\mathbb{Z}_2^A \times \mathbb{Z}_2^B$. When $J > 0$, the ground state is a spin singlet; when $J < 0$, as long as there are terms that reduce the symmetry to $\mathbb{Z}_2^A \times \mathbb{Z}_2^B$, the ground state is in general nondegenerate, hence any spin correlation along the temporal direction would still be short ranged. 

As was understood before, multiple copies of fermionic topological insulators (TI) and topological superconductors (TSC) of fermions may be mapped to bosonic SPT states under interaction~\cite{Metlitski1406.3032,wangsenthil2014, bridge,youxu2014}. For example, four copies of Kitaev's chains of Majorana fermions, or two copies of the spinless Su–Schrieffer–Heeger (SSH) models of complex fermions~\cite{SSH}, can be mapped to the Haldane phase with different defining symmetries. Our evaluation above also implies that, under general decoherences on fermion bilinear operators that respect the symmetry of the system after averaging over all quantum trajectories, the classification of the fermionic TIs and TSCs would collapse, analogous to the collapse of the classification of TIs and TSSs under short-range interactions~\cite{fidkowski1,fidkowski2}). Similar behavior of TIs and TSCs under decoherence in higher dimensions is also expected, and it was noticed that the averaged symmetry could also lead to the collapse of classifications of TIs~\cite{MaWang2022}. We will defer a more complete discussion of decohered TIs and TSCs to future work.

Now let's consider another type of interaction: 
\begin{align}\label{int2} 
    \mathcal{S}^{\textrm{int}}_{1d,2}[\vect{n}(x), \vect{n}'(x)] \sim \int dx \ u \left( n_z (x)n'_z(x) \right). 
\end{align}
Here only the $z$ components of the two $\vect{n}$ vectors are coupled, which corresponds to introducing decoherence on $\mathbb{Z}_2^B$ but not $\mathbb{Z}_2^A$, as now the density matrix is only invariant under one simultaneous $\mathbb{Z}_2^B$ transformation on $n_z(x)$ and $n_z'(x)$, but it is still invariant under two separate $\mathbb{Z}_2^A$ symmetry transformations on $\vect{n}$ and $\vect{n}'$. Then after the space-time rotation, the strange correlator calculation is mapped to a temporal spin-spin correlation (either \eqnref{scItemporal} or \eqnref{scIItemporal}) of the following two interacting spins: 
\begin{align}
    H^{\textrm{int}}_{0d,2} \sim J S_z S'_z.
\end{align}
Then the ground state of the system would be a doublet, rather than a singlet. For example, when $J < 0$, the ground states of the two-spin system are 
\begin{align}
    |\mathrm{GS}_1\rangle &= |S_z = + 1/2, \ S'_z = +1/2\rangle,\nonumber \\ |\mathrm{GS}_2\rangle &= |S_z = -1/2, \ S'_z = -1/2\rangle.
\end{align}
The type-I strange correlator $C^\mathrm{I}_{xx}(x)$ defined in \eqnref{eq:strange_typeI} evaluated as \eqnref{scItemporal} would be short ranged, as a single $S_x$ operator does not connect these two states within the doublet; { however, the type-II strange correlator $C^{\mathrm{II}}_{xx}(x)$ is still long-ranged, as the operator $S_x S'_x$ can connect the two states within the doublet}. Both $C^\mathrm{I}_{zz}$ and $C^{\mathrm{II}}_{zz}$ are long-ranged, as both the ground states are eigenstates of $S_z$ and $S'_z$.

It was concluded before that certain edge state of the SPT state with $\mathbb{Z}_2 \times \mathbb{Z}^\text{avg}_2$ symmetry might be trivialized by edge symmetry breaking (though the SPT state was still considered nontrivial)~\cite{MaWang2022}, where $\mathbb{Z}^\text{avg}_2$ is the average symmetry that only exists after disorder average. Our calculation indicates that the boundary state, which is related to the type-I strange correlator, may not be the best diagnosis for systems under decoherence; instead, the type-II strange correlator serves as another method of characterizing mixed-state SPT phases. The type-II strange correlator is to some extent analogous to the Edwards-Anderson correlator of spin glass systems. Indeed, as remarked in the numerical lattice model calculation, for each quantum trajectory the type-I strange correlator fluctuates in its sign while magnitude stays robust, signaling the presence of nontrivial correlation. Although the type-I strange correlator vanishes over averaging, the squared order parameter, i.e., the type-II strange correlator, still captures the nontrivial information from the SPT state.

\subsection{$2d$ SPT states}

Now let us consider the wave function of a class of $2d$ SPT states, which is derived from the bulk NLSM following the discussions in Ref.~\cite{sptwf,YouXu2013}: 
\begin{align}
 &|\Psi \rangle  \sim \int D[\vect{n}(\mathbf{x})]
\exp\left( - \mathcal{S}[\vect{n}(\mathbf{x})] \right)
|\vect{n}(\mathbf{x}) \rangle \nonumber \\
& \mathcal{S}  = \int d^2x \
\frac{1}{g} (\partial \vect{n}(\mathbf{x}))^2 +
\mathrm{WZW}[\vect{n}(\mathbf{x})] \nonumber \\
&\mathrm{WZW}[\vect{n}(\mathbf{x})] = \frac{2\pi \ii}{\Omega_3} \int_0^1 du \int d^2x \nonumber \\
& \hspace{40pt}\epsilon_{abcd} \tilde{n}^a
\partial_u \tilde{n}^b \partial_x \tilde{n}^c \partial_y \tilde{n}^c.
\label{2dnlsm} \end{align}
The pure density matrix is analogous to \eqnref{puredm}. Similar to the 1d case, this formalism can describe SPT phases with symmetries $\O(4)$, $\U(1) \times \U(1)$, $\SO(3) \rtimes \mathbb{Z}_2$, $\SO(3) \times \mathbb{Z}_2^T$, $\mathbb{Z}_2 \times \mathbb{Z}_2$, or even just one $\U(1)$ or $\mathbb{Z}_2$ symmetry, etc. 

Let us start with the case with the maximum $\O(4)$ symmetry. We can introduce some weak decoherence on $\vect{n}$, and consider the mixed density matrix \eqnref{mixeddm}. We start with an interaction term analogous to \eqnref{int1d1}: \beqn \cS^{\textrm{int}}_{2d,1} = \int d^2x \ u \left( \vect{n}(\mathbf{x}) \cdot \vect{n}'(\mathbf{x}) \right). \eeqn This interaction respects the simultaneous left and right $\O(4)$ symmetry (which act on $\vect{n}(\mathbf{x})$ and $\vect{n}'(\mathbf{x})$ simultaneously). Like the $1d$ case, the decoherence is supposed to drive the system into a mixed state with enhanced weight on the diagonal configurations, hence $u$ is most naturally negative. But regardless of the sign of $u$, this interaction drives the computation of both type-I and type-II strange correlators into a field theory calculation of correlation functions of a $(2+0)d$ or $(1+1)d$ nonlinear Sigma model (NLSM) without any WZW term. Both type-I and type-II strange correlators should be short-ranged, as an NLSM in $(2+0)d$ without any topological term will flow to the disordered phase with only short-range correlations. 

Now let us consider a $2d$ bosonic SPT state with symmetry $\U(1)^A \times \U(1)^B$. This is an SPT phase considered in Ref.~\cite{levinsenthil}. The two $\U(1)$ symmetries act on the four-component vector $\vect{n}$ as the following: 
\begin{align}
    \U(1)^A &: (n_1 + \ii n_2) \rightarrow e^{\ii \theta} (n_1 + \ii n_2), \nonumber \\ \U(1)^B &: (n_3 + \ii n_4) \rightarrow e^{\ii \phi} (n_3 + \ii n_4). \label{u1ab} 
\end{align}
Let us turn on the following interaction, which corresponds to introducing decoherence on $\U(1)^B$ charges: 
\begin{align}
    \cS^{\textrm{int}}_{2d,2}[\vect{n}(\mathbf{x}), \vect{n}'(\mathbf{x})] \sim \int d^2x \ \sum_{a = 3}^4 u ( n_a (\mathbf{x}) n'_a(\mathbf{x}))
\end{align}
This interaction can be conveniently analyzed through Abelian bosonization of the $(1+1)d$ NLSM. Under Abelian bosonization, the NLSM of $\vect{n}(\bf{x})$ with the WZW term corresponds to the following $(1+1)d$ Lagrangian, and its dual: 
\begin{align}
    \label{abelian}
\cL &= \frac{1}{4\pi K} \left( (\partial_\tau \theta)^2 + v^2 (\partial_x \theta)^2 \right) , \nonumber \\ 
\cL_d &= \frac{K}{4\pi}\left( (\partial_\tau \phi)^2 + v^2 (\partial_x \phi)^2 \right),
\end{align}
where we require local excitations $e^{i\phi}$ and $e^{i\theta}$ to commute with each other (mutual bosonic) when they are not on the same space position.
The four-component vector $\vect{n}$ has the following schematic representation in the Abelian bosonized formalism: 
\beqn 
\vect{n} \sim \left(\cos\theta, \sin\theta, \cos \phi, \sin \phi \right). 
\eeqn 
The computation of the strange correlator, especially type-II, requires making two copies of the system of \eqnref{abelian}, and now the interaction term $\cS^{\textrm{int}}_{2d,2}$ becomes the following coupling in the Abelian bosonization language: 
\beqn \label{eq:interaction}
H^{\textrm{int}}_{1d,2} \sim \int dx \ \alpha
\cos( \phi - \phi'). 
\eeqn

We evaluate the following type-I and type-II strange correlators for the operator $\Phi \sim n_1 - \ii n_2 \sim e^{-\ii \theta}$ charged under $\U(1)^A$:
\beqn
\begin{split} 
C^{\mathrm{I}}(\mathbf{x}) &= \frac{\mathrm{tr}\left( \Phi (0) \Phi^\ast(\mathbf{x}) \rho^{D}_{\mathrm{spt}} \rho^\vdagger_0 \right)}{\mathrm{tr}\left(\rho^{D}_{\mathrm{spt}} \rho^\vdagger_0 \right)}\\
&\sim \langle e^{- \ii \theta(0,0)} \ e^{+ \ii \theta(x,\tau)} \rangle,\\
C^{\mathrm{II}}(\mathbf{x}) &= \frac{\mathrm{tr}\left( \Phi (0) \Phi^\ast (\mathbf{x})\rho^{D}_{\mathrm{spt}} \Phi (0) \Phi^\ast (\mathbf{x}) \rho_0 \right)}{\mathrm{tr}\left(\rho^{D}_{\mathrm{spt}} \rho^\vdagger_0 \right)} \\ 
&\sim \langle e^{- \ii ( \theta(0,0) + \theta'(0,0))} \ e^{+ \ii (\theta(x,\tau) + \theta'(x,\tau))} \rangle,
\end{split} \label{sc2d} 
\eeqn 
where the expectation values are taken with respect to the two copies of Luttinger liquids coupled by \eqnref{eq:interaction}.

As the scaling dimension of $e^{i \phi}$ is given by $\frac{1}{2K}$, the coupling $H^{\textrm{int}}_{1d,2}$ is relevant (irrelevant) when $K\,{>}\,K_c$ ($K\,{<}\,K_c$) with $K_c = 1/2$. When $K\,{>}\,K_c$, $i.e.$ $H^{\textrm{int}}_{1d,2}$ is relevant, it will gap out the channel $\theta_- = (\theta - \theta')/\sqrt{2}$, $\phi_- = (\phi - \phi')/\sqrt{2}$, and the type-I strange correlator becomes short-ranged. Hence by tuning $K$, there is a phase transition signified by the behavior of the type-I strange correlator. Since the Luttinger parameter $K$ can receive renormalization from $\alpha$ in \eqnref{eq:interaction}, this phase transition of the type-I strange correlator can also be driven by the strength of interaction. But the $\theta_+ = (\theta + \theta')/\sqrt{2}$ and $\phi_+ = (\phi + \phi')/\sqrt{2}$ channel remains a gapless CFT, regardless of the fate of $H^{\textrm{int}}_{1d,2}$ under renormalization group. Since the type-II strange correlator in \eqnref{sc2d} only involves the symmetric linear combinations, it always decays with a power law. Here once again the type-II strange correlator captures the ``memory'' of the pure state SPT wave function. The type-II strange correlator being nontrivial exemplifies the fact that there are still some non-trivial topological features in the system even though we reduce the doubled $\U(1)^B$ symmetries down to their diagonal subgroup.

\subsection{Bosonic Integer Quantum Hall state}

We would also like to discuss the $2d$ bosonic integer quantum Hall (bIQH) state discussed in Ref.~\cite{levinsenthil}. The bIQH state can be obtained from the same NLSM description \eqnref{2dnlsm}, and reducing the $\U(1)^A \times \U(1)^B$ symmetry in \eqnref{u1ab} to one diagonal $\U(1)$ symmetry. The edge state of the bIQH state is nonchiral, whose left-moving modes ($\phi+\theta$) carry the $\U(1)$ charge, while the right-moving ($\phi-\theta$) modes are neutral. The Hall conductivity of the bIQH state must be an even integer due to the bosonic nature of the underlying particles. 

When we consider the density matrix of the bIQH state, the $\U(1)$ symmetry again gets doubled. Assume we choose $e^{\ii \theta}$ and $e^{\ii \phi}$ to carry a positive charge under $\U(1)$. Then, $e^{-\ii \theta'}$ and $e^{\ii \phi'}$ carry a positive charge under $\U'(1)$. This is because the theory of $\theta'$ and $\phi'$ must be the conjugate of the theory of $\theta$ and $\phi$~\footnote{More precisely, as discussed in \appref{app:Choi}, the SPT wavefunction for the other copy must be conjugate of the original SPT wavefunction.}. We consider decoherence that breaks the $\U(1) \times \U(1)'$ symmetry down to their diagonal subgroup; this effect at the lowest order is mapped to the following interaction between the two copies of the system, after space-time rotation: 
\begin{align}
    H^{\textrm{int}}_{1d,3} = \int dx \ \alpha\cos(\phi - \phi') + \beta\cos(\theta + \theta').
\end{align}
Since two terms in $H^{\textrm{int}}_{1d,3}$ have scaling dimensions $K$ and $1/K$ respectively, the Luttinger parameter can be chosen such that both $\pm$ channels of \eqnref{abelian} are gapped out. This implies that both type-I and II strange correlators of the bIQH state can be short-ranged under the decoherence which only preserves the diagonal symmetry $\U(1)$. 

Here we note that a similar analysis also applies to the fermionic integer quantum Hall states, whose boundary states become chiral bosons after Abelian bosonization. A full analysis of fermionic topological insulators under decoherence will be presented in another work. 

One can also consider the physical boundary state of $2d$ SPT phases under decoherence, by deriving the wave function of the $1d$ boundary state of the SPT phase and turning on decoherence on the boundary wave function. This analysis would be similar to the Luttinger liquid under weak measurement discussed in Ref.~\cite{Garratt2022}. 

\section{Doubled System} \label{sec:doubled}


In this section, we develop a more systematic understanding of the SPT ensemble, where we use Choi-Jamiołkowski isomorphism~\cite{JAMIOLKOWSKI1972, CHOI1975} to faithfully map a mixed state density matrix into a pure state defined in the doubled Hilbert space.
In doing so, we illustrate the example of a mixed state obtained by applying a decohering quantum channel to a pure SPT state, showing that such a mixed state can retain its topological feature. This idea can be applied to any mixed state representing a symmetric ensemble of quantum states, and hence an SPT ensemble can be generally defined as a density matrix that corresponds to a pure SPT state in the doubled Hilbert space. We will show that strange correlators discussed in the previous section become observables that can diagnose some features of a pure SPT state in the doubled Hilbert space. 

\subsection{Choi Representation of Decohered SPT States} \label{sec:doubleH}

Let $\rho_\spt$ be the density matrix of the pure SPT state with $G\,{=}\,G_A\,{\times}\,G_B$ symmetry. By using the Choi-Jamiołkowski isomorphism, we can represent an operator as a state in the following way
\begin{equation}
    \Vert \rho \rangle \hspace{-2pt} \rangle  \equiv \frac{1}{\sqrt{ \textrm{Dim}[\rho] }} \sum_i |i \rangle \otimes \rho |i\rangle 
\end{equation}
Accordingly, the pure state density matrix can be represented as two copies of independent SPT states:
\begin{equation}
    \Vert \rho_\spt \rAngle = |\Psi^* \rangle_u \otimes | \Psi \rangle_l = | \Psi^*_u, \Psi_l \rangle.
\end{equation}
where $u$ and $l$ denote the upper and lower copies respectively. The Choi state is defined on the doubled Hilbert space as ${\cal H}_d \equiv {\cal H}_u \otimes {\cal H}_l$. 
Here, the star superscript is to denote that its amplitude is complex-conjugated relative to the original wavefunction. In $\cH_d$, the symmetry group is also doubled as $(G^{u}_A\,{\times}\,G^u_B)\,{\times}\,(G^l_A\,{\times}\,G^l_B)$, and the Choi state is the SPT state under the doubled symmetry group. See \appref{app:Choi} for more details. Similarly, the Choi state of a trivial disordered state density matrix is given as
\begin{equation}
    \Vert \rho_0 \rAngle  = |\Omega^* \rangle_u \otimes |\Omega \rangle_l \equiv | \tilde{\Omega} \rangle 
\end{equation}
which is a trivial disordered state in $\cH_d$.

Under the Choi isomorphism, the decoherence quantum channel $\cE$ in \eqnref{eq:noise_Kraus} maps into an operator $\hat{\cE}$ acting  in $\cH_d$ as the following:
\begin{align}
    \hat{\cE} = \sum_{m} K^*_{m,u} \otimes K_{m,l}
\end{align}
where $K_m$ is a Kraus operator. For example, the aforementioned local $Z$ noise channel under the Choi mapping would be expressed as
\begin{equation} \label{eq:noise_model1}
    \hat{\cE} = \prod_i \big[ (1-p) \mathbb{I}_{i,u} \otimes \mathbb{I}_{i,l} + p Z_{i,u} \otimes Z_{i,l} \big].
\end{equation}
Then, the decohered SPT state can be represented as 
\begin{equation}
    |\tilde{\Psi} \rangle \equiv \Vert \cE[\rho_\text{spt}] \rAngle = \hat{\cE} | \Psi_u, \Psi_l \rangle 
\end{equation}
Although $\hat{\cE}$ is not a unitary map that preserves the norm of the state, it is nevertheless a positive semi-definite map for the Choi states. Therefore, $|\tilde{\Psi}\rangle$ is a valid state in the doubled Hilbert space whose norm is given by the purity of the decohered density matrix, $\textrm{tr}(\rho^D_\spt \rho^D_\spt) > 0$. 

The Choi representation of the mixed state density matrix $|\tilde{\Psi} \rangle$ can be shown to be the ground state of a certain parent Hamiltonian in $\cH_d$, which is perturbed from the doubled SPT Hamiltonian. Starting from the pure state limit, we consider a pure SPT state $\ket{\Psi}$ as the ground state of a parent Hamiltonian $H_\text{spt}$ in the standard Hilbert space. With some constant energy shift, it is always possible to make $H_\text{spt}$ a positive semidefinite Hermitian operator, such that the ground state has exactly zero energy, i.e. $H_\text{spt}\ket{\Psi}=0$. Correspondingly, in the doubled Hilbert space, the Choi representation of the SPT density matrix $\Ket{\rho_\text{spt}}=\ket{\Psi^*}_u\otimes\ket{\Psi}_l$ should be the ground state of the following double Hamiltonian
\begin{equation}
H^d_\text{spt}=H_{\text{spt},u}^*\otimes \mathbb{I}_l +  \mathbb{I}_u\otimes H_{\text{spt},l},
\end{equation}
such that $H^d_\text{spt}\Ket{\rho_\text{spt}}=0$. The double Hamiltonian $H^d_\text{spt}$ will inherit the positive semidefinite property of the single Hamiltonian $H_\text{spt}$, such that it admits a Cholesky decomposition as 
\begin{equation}
    H^d_\text{spt}=\hat{\cA}^\dagger\hat{\cA},
\end{equation}
with $\hat{\cA}$ being some generic (possibly non-Hermitian) operator in the doubled Hilbert space. The fact that $\Ket{\rho_\text{spt}}$ is a zero-energy eigenstate of $H^d_\text{spt}$ implies 
\begin{equation}
\Bra{\rho_\text{spt}}H^d_\text{spt}\Ket{\rho_\text{spt}}=\Bra{\rho_\text{spt}}\hat{\cA}^\dagger\hat{\cA}\Ket{\rho_\text{spt}}=0,
\end{equation}
therefore $\hat{\cA}\Ket{\rho_\text{spt}}=0$ must be a zero vector, i.e.~the ground state $\Ket{\rho_\text{spt}}$ should be annihilated by the Cholesky operator $\hat{\cA}$.

In our setup, the decohered SPT state $\rho_\text{spt}^D\,{=}\,\cE[\rho_\text{spt}]$ is always given by sending the pure SPT state $\rho_\text{spt}\,{=}\,|\Psi \rangle \langle \Psi |$ through a decoherence channel $\cE$. In the Choi representation, this can be written as $\Ket{\rho_\text{spt}^D}\,{=}\,\hat{\cE}\Ket{\rho_\text{spt}}$. We claim that the state $\Ket{\rho_\text{spt}^D}$ must be the ground state of the following deformed double Hamiltonian
\begin{equation}\label{eq:HsptD_nonlocal}
    H^d_\text{spt}=(\hat{\cE}\hat{\cA}\hat{\cE}^{-1})^\dagger (\hat{\cE}\hat{\cA}\hat{\cE}^{-1}),
\end{equation}
which is still a Hermitian positive semidefinite operator in the doubled Hilbert space. To see this, we apply  $H^d_\text{spt}$ on the state $\Ket{\rho_\text{spt}^D}$ and obtain $H^d_\text{spt}\Ket{\rho_\text{spt}^D}
=(\hat{\cE}\hat{\cA}\hat{\cE}^{-1})^\dagger \hat{\cE}\hat{\cA}\Ket{\rho_\text{spt}}=0$, which proves our claim. In this way, given the original SPT Hamiltonian $H_\text{spt}$ and the decoherence channel $\cE$, we can in principle construct the parent Hamiltonian $H^d_\text{spt}$ that stabilizes the decohered SPT state $\Ket{\rho_\text{spt}^D}$ as its unique ground state.

However, one concern is that the Hamiltonian constructed in \eqnref{eq:HsptD_nonlocal} may not be a local Hamiltonian. Nevertheless, if the original SPT Hamiltonian $H_\text{spt}$ is made of local commuting projectors (e.g.~a stabilizer Hamiltonian), it is possible to define a set of local Cholesky operators $\hat{\cA}_i$ (using local projectors) such that $H^d_\text{spt}=\sum_{i}\hat{\cA}_i^\dagger \hat{\cA}_i$, then the same construction leads to a local Hamiltonian
$H_\text{spt}^d=\sum_i(\hat{\cE}\hat{\cA}_i\hat{\cE}^{-1})^\dagger (\hat{\cE}\hat{\cA}_i\hat{\cE}^{-1})$~\cite{WITTEN1982, Wouters2021, pivot}.

We now apply this general construction principle to $1d$ and $2d$ cluster state examples discussed previously.

\begin{figure}[!t]
    \centering
    \includegraphics[width = 0.47 \textwidth]{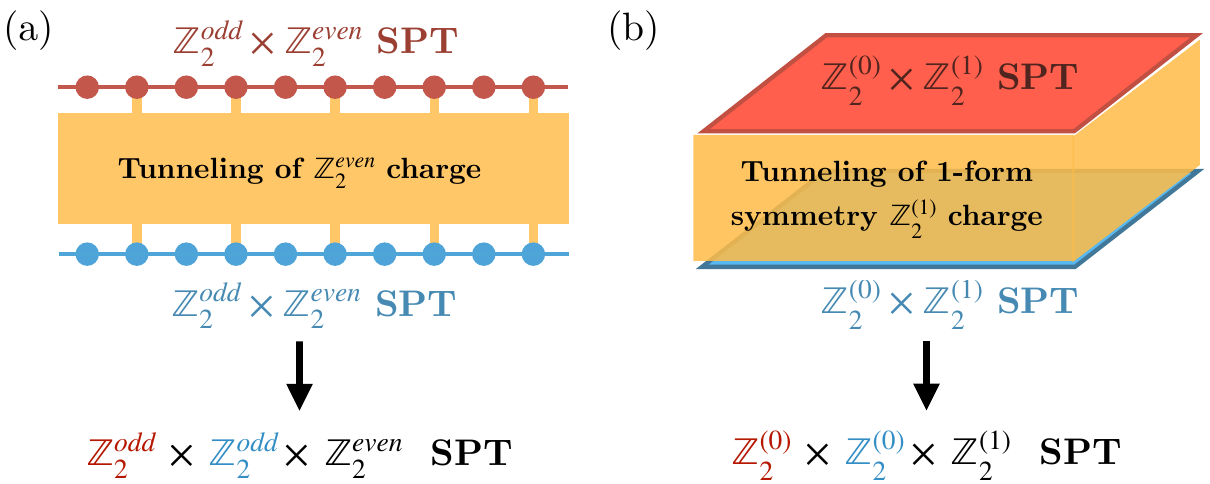}
    \caption{\label{fig:doubled} {\bf SPT structure of Choi states in the doubled Hilbert space.} (a) 1d $\mathbb{Z}_2^\todd \times \mathbb{Z}_2^\teven$ symmetric cluster state density matrix and (b) 2d $\mathbb{Z}_2^{0} \times \mathbb{Z}_2^{1}$ symmetric cluster state density matrix under the Choi Isomorphism. The decoherence of the symmetry $G_B$ maps to the coupling between $G_B^u$ and $G_B^l$ charges in two layers. Accordingly, two respective symmetries become identified, and the SPT classification reduces.   }
    
\end{figure}

\subsubsection{1d cluster state}

For the $1d$ cluster state, the SPT projector Hamiltonian can be written as 
\begin{equation}
    H_\text{spt}=
    \sum_{i} \hat{A}_i^2, \quad \hat{A}_i \equiv \frac{1-Z_{i-1}X_iZ_{i+1}}{2}.
\end{equation}
Consider the decoherence model in \eqnref{eq:noise_model1}, which has the following convenient form
\begin{align} \label{eq:noise_model}
     \hat{\cE} &= \prod_m  (1-2p)^{1/2} e^{\tau Z_{2m,u} Z_{2m,l}}, \quad \tanh \tau = \frac{p}{1-p}
\end{align}
Then, we see that
\begin{align}
    \hat{\cE}\hat{\cA}_{2n} \hat{\cE}^{-1} = \frac{1}{2} \big( 1 - e^{2\tau Z_{2n,u} Z_{2n,l}} Z_{2n-1} X_{2n} Z_{2n+1} \big)
\end{align}
while $\hat{\cE} \hat{\cA}_{2n-1} \hat{\cE}^{-1}\,{=}\,\hat{\cA}_{2n-1}$. As depicted in \figref{fig:doubled}(a), the parent Hamiltonian $H_\text{spt}^d$ for the decohered SPT state takes the form of 
\begin{equation}\label{eq: HsptD}
    H_\text{spt}^d= H_{\text{spt},u}^d+H_{\text{spt},l}^d+H_{\textrm{int}}^d,
\end{equation}
where the upper layer Hamiltonian reads
\begin{align}
    H_{\text{spt},u}^d &=  \sum_m \frac{\cosh^2 2\tau  - \cosh 2\tau Z_{2m-1} X_{2m} Z_{2m+1}}{2} \nonumber \\
    &\,+ \sum_m \frac{1  - Z_{2m} X_{2m+1} Z_{2m+2}}{2},
\end{align}
and the lower layer Hamiltonian $H_{\text{spt},l}^d$ is essentially the same as $H_{\text{spt},u}^d$ with all the label $u$ replaced by the label $l$, together with the interlayer coupling
\begin{equation} \label{eq:coupling1D}
H_{\textrm{int}}^d=-\frac{\sinh 4\tau}{2}\sum_{m}Z_{2m,u}Z_{2m,l}.
\end{equation}
The coupling vanishes in the pure state limit when the decoherence strength $p\to 0$ ($\tau \to 0$), and diverges in the strong decoherence limit as $p\to 1/2$ (which corresponds to measuring $Z_{2m}$ operators projectively and forgetting the measurement outcomes).

This parent Hamiltonian $H_\text{spt}^d$ in \eqnref{eq: HsptD} explicitly shows that the decohered SPT state $\Ket{\rho_\text{spt}^D}$ can be interpreted as two identical layers of $\mathbb{Z}_2^\todd\times\mathbb{Z}_2^\teven$ SPT states coupled together by the interlayer $\mathbb{Z}_2^\teven$ charge tunneling (i.e.~the ferromagnetic interlayer $ZZ$ coupling on every even site). The resulting state is a nontrivial SPT state protected by the $\mathbb{Z}_2^{\todd,u}\times\mathbb{Z}_2^{\todd,l}\times\mathbb{Z}_2^{\teven}$ symmetry in $\cH_d$. This SPT is characterized by the following nonlocal order parameter $S_\textrm{odd}$:
\begin{align}
    S_\textrm{odd}(2r) :=  Z_{1,u} Z_{1,l} \Big(\prod_{i=1}^r X_{2i,u} X_{2i,l} \Big) Z_{2r+1,u} Z_{2r+1,l} 
\end{align}
Note that $S_{\textrm{odd}}(2r)$ commute with the couping \eqnref{eq:coupling1D}; thus, with respect to $\Vert \rho^D_\spt \rAngle$ the expectation value $\langle  S_\textrm{odd}(2r) \rangle$  always takes a unit value for all $r$ and $p$. This lattice model analysis is in full agreement with the field theory understanding in \secref{sec:1d field theory}.

\subsubsection{2d cluster state}

For the $2d$ cluster state under the local $Z$ noise on edge qubits described by \eqnref{eq:noise_model}, the Choi state of the decohered density matrix can be shown to be the groundstate of the Hamiltonian in the form \eqnref{eq: HsptD}, where
\begin{align} \label{eq:2dSPT_couple}
    H_{\text{spt},u}^d &= - \cosh 2 \tau \sum_e \bm{X}_{e} \prod_{v \in \rd e} Z_{v} - \sum_v X_v \prod_{e \in \dd v} \bm{Z}_e  \nonumber \\
    H_{\textrm{int}}^d&= - \sinh 4\tau  \sum_{e} \bm{Z}_{e,u}\bm{Z}_{e,l}.
\end{align}
In this doubled system, the coupling  $H_\textrm{int}^d$ breaks two one-form symmetries into their diagonal subgroup, and the remaining global symmetry in the doubled Hilbert space would be $\mathbb{Z}_2^{(0),u} \times \mathbb{Z}_2^{(0),l} \times \mathbb{Z}_2^{(1)}$  as illustrated in \figref{fig:doubled}(b).

In particular, the $\mathbb{Z}_2^{(1)}$ one-form symmetry \cite{Kapustin2013H1309.4721,Gaiotto2015G1412.5148,McGreevy2023G2204.03045} is implemented by the operator $\prod_{e\in \partial\mathcal{A}}\bm{X}_{e,u}\bm{X}_{e,l}$, where $\partial\mathcal{A}$ denotes any closed loop along the links on the lattice. 
The doubled Hamiltonian is invariant under the associated one-form symmetry for any choice of a closed-loop $\partial\mathcal{A}$.

Microscopically, this $\mathbb{Z}_2^{(1)}$ is the only explicit 1-form symmetry, which is a diagonal subgroup of two 1-form symmetries. However, note that 1-form symmetries can be often emergent, exhibiting robustness against perturbations that breaks the symmetries~\cite{Tupitsyn2010, Nahum2021}. Indeed, for $p\,{\ll}\,1$, two emergent one-form symmetries still exist, providing a bulk anomaly in-flow such that if one imagines a fictitious boundary of the doubled system, the boundary would exhibit spontaneous symmetry breaking of individual $\mathbb{Z}_2^{(0),u}$ or $\mathbb{Z}_2^{(0),l}$ symmetries. This, in turn, implies that the type-I strange correlator for $G_A^u \equiv \mathbb{Z}_2^{(0),u}$ charged operators must be nontrivial in the doubled (pure) state for small $p$. In other words,
\begin{align}
    C^\textrm{I}_{A_u}(v,v')  &\equiv \frac{     \langle \tilde{\Omega} | \, Z_v Z_{v'} \otimes \mathbb{I}  \,| \tilde{\Psi} \rangle    }{     \langle \tilde{\Omega} | \tilde{\Psi} \rangle   }
\end{align}
would be nontrivial. Since this expression is exactly the type-I strange correlator for the original decohered mixed state, the nontrivial type-I strange correlator under decoherence in this model can be understood in terms of the stability of the emergent one-form symmetries in the doubled system.

At a more intuitive level, the robustness of an SPT state protected by higher-form symmetries under the local decoherence can be thought of as a consequence of extensively many conserved quantities, which is in contrast with the single globally conserved charge of a 0-form symmetry. Accordingly, some decoherence in local charge configuration cannot destroy the macroscopic stability of emergent higher-form symmetries, and the corresponding topological response acquires noise resilience.

\subsection{SPT ensemble}

As we have examined for a few important examples, Choi isomorphism provides a way to understand SPT characteristics of a given mixed state density matrix by ``faithfully'' mapping it into a pure state. This pure state in the doubled state would enjoy an enlarged symmetry and an anomaly protected by this enlarged symmetry, thus being a non-trivial SPT state in the doubled Hilbert space. On the other hand, if a mixed state is decohered too much, the corresponding pure state would be a trivial state in the doubled Hilbert space. We denote a mixed state (ensemble of states) that maps to a non-trivial SPT state in the doubled Hilbert space as an SPT ensemble.

In general, for a given pure state in the doubled Hilbert space, there are various diagnostics to identify its non-trivial SPT features; a conventional strange correlator is one of them. Interestingly, generalized strange correlators of a mixed state density matrix correspond to diagnostics of a pure state SPT in the doubled Hilbert space, which is a conventional strange correlator as the following:
\begin{align} \label{eq:boundary}
    C^\textrm{I}_{A} &= \frac{     \langle \tilde{\Omega} | \, (O^A_i O^A_j)_u \otimes \mathbb{I}_l  \,| \tilde{\Psi} \rangle    }{     \langle \tilde{\Omega} | \tilde{\Psi} \rangle   } = C_{A_u} \\
    C^\textrm{II}_A &= \frac{     \langle \tilde{\Omega} | \, (O^A_i O^A_j)_u \otimes (O^A_i O^A_j)_l  \,| \tilde{\Psi} \rangle    }{     \langle \tilde{\Omega} | \tilde{\Psi} \rangle   } = C_{A_u\cdot A_l},
\end{align}
where $C$ without superscript denotes that it is a conventional strange correlator defined for a pure state (in the doubled Hilbert space), and the subscript $A_u \cdot A_l$ means that the strange correlator is for the operator charged under both $G_A^u$ and $G_A^l$. 

As the type-II strange correlator of the mixed state $\rho_\spt^D$ for $G_A$ is the conventional strange correlator of the Choi state $\Vert \rho_\spt^D \rAngle$ for the operator charged under both $G_A^u$ and $G_A^l$,
it naturally probes the boundary mixed anomaly of the $G_A^u \times G_A^l \times G_B$ SPT phase, as in \figref{fig:summary}(d). In \secref{sec:nonlocal}, we show that a strange correlator provides a mathematically equivalent way of measuring the presence of a boundary mixed anomaly in the pure state. Since the Choi operator of the decoherence $\hat{\cE}$ acts like a shallow quantum circuit acting on the doubled SPT state that respects the symmetry, as long as the decoherence is not too strong, the $G_A^u \times G_A^l \times G_B$ SPT order should be robust. Therefore, this Choi isomorphism naturally provides a mechanism for the stability of the type-II strange correlator under selective decoherence, and further demonstrates how this is tied to the stability of an SPT ensemble.

\begin{figure}[!t]
    \centering
    \includegraphics[width = 0.43 \textwidth]{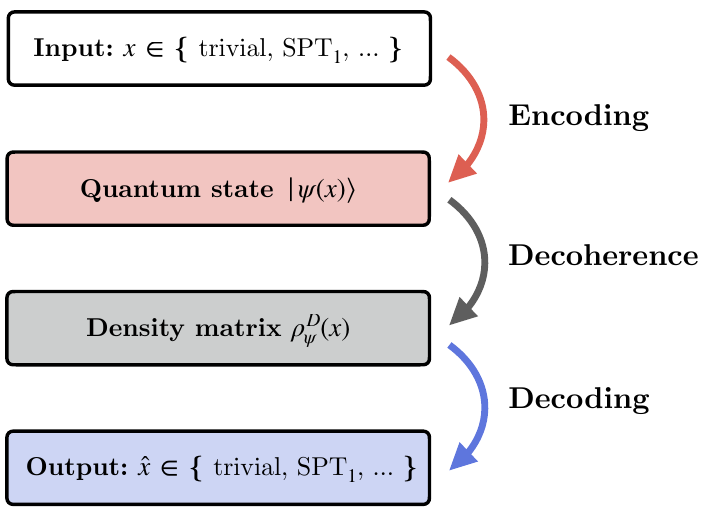}
    \caption{\label{fig:information} {\bf Encoding, Decoherence, and Decoding of Information using quantum states}. After the encoding step, a label $x$ is encoded in quantum correlations of a prepared state $|\psi(x)\rangle$. The state undergoes decoherence, becoming a mixed state $\rho^D(x)$. Finally, a receiver measures this mixed state to learn about the label $x$ (assuming the state is repeatedly prepared). After enough measurements, the receiver may generate an output $\hat{x}$ which is the estimator of the label $x$. The decoding is successful if $\hat{x} = x$.
    } 
\end{figure}

\subsection{Physical Implications}

A nontrivial type-II strange correlator has profound implications. First of all, the type-II strange correlator being nontrivial means that, in the doubled system, even though the symmetry is reduced due to decoherence, there is still a mixed topological response. Also, the type-II strange correlator can be interpreted as an information-theoretic quantity that allows one to identify what the underlying state is for a given mixed-state density matrix. As illustrated in \figref{fig:information}, one may imagine a procedure where the \emph{label} information $x$ is encoded as the quantum circuit that can repeatedly generate a certain quantum state. For example, given a label $x\,{=}\,\textrm{SPT}$, a corresponding quantum circuit that prepares a certain SPT state can be constructed, whose output quantum state is sent to an observer. While doing so, decoherence occurs for various reasons, such as imperfection in the state preparation or transmission noise from environments, converting the pure quantum state into a mixed state density matrix. Finally, the observer repeatedly measures the mixed state to learn about the input label $x$. This last step is formally denoted as \emph{decoding}. 

Whether one can accurately identify the label or not crucially depends on the decoding strategy. For example, if there is no decoherence, decoding would be straightforward; one can simply measure the type-I strange correlator, which as we argued can be viewed as a non-local SPT order parameter for a given pure state density matrix, and its value would immediately allow one to tell the decoded label $\hat{x}$ (See \secref{sec:measure} for more precise protocol). Such a decoding strategy would also be efficient in the sense that the required number of measurements would at most scale polynomially with the system size. Although the type-I strange correlator may become short-ranged with strong enough decoherence, at weak decoherence the type-I strange correlator may remain nontrivial. Hence for weak decoherence, the type-I strange correlator can still tell us whether the underlying state was a trivial disordered state or a nontrivial SPT state.

But with strong decoherence, we have shown that there are examples where the type-I strange correlator becomes short-ranged, while the type-II strange correlator remains nontrivial. Indeed, the robustness of the type-II strange correlator implies that there must be a method to distinguish between the decohered mixed states originating from an SPT state and those originating from a trivial disordered state. Without referring to any decoding strategy, the fact that the decohered trivial state and decohered SPT state are distinguished by the type-II strange correlator implies the existence of a fundamental information-theoretic distinction between these two labels. Although the corresponding decoding strategy may not be efficient, the type-II strange correlator behavior provides a fundamental distinction between two classes of density matrices. Finally, when both the type-I and type-II strange correlators are trivial, one cannot tell whether the underlying state is an SPT or trivial state based on any feature that involves the mixed state density matrix. Therefore, behaviors of the type-I and type-II strange correlators allow us to distinguish three different regimes of the mixed state density matrix under decoherence as illustrated in \figref{fig:summary}(e).

A nontrivial Type-I strange correlator also implies that the physical boundary cannot be trivialized (though its exact behavior, i.e. whether it is gapless or develops long-range order, depends on the detailed boundary Hamiltonian); a nontrivial type-II strange correlator implies that the boundary in the doubled Hilbert space cannot be trivialized, as the type-II strange correlator is type-I in the doubled Hilbert space; but in order to observe the nontrivial boundary effect that corresponds to the type-II strange correlator in the doubled Hilbert space, we need to observe quantities nonlinear with density matrix at the boundary. 


Here we comment on the efficiency of measuring the type-II strange correlator or an observable that can measure an ordering in the doubled Hilbert space. 
Since a quantity defined in the doubled Hilbert space would require the contraction of two Choi states (bra and ket), the quantity is challenging to measure in practice. This is because such a quantity would require an additional normalization by the overlap of two Choi states; for example, the numerator and denominator of the type-II strange correlator should be evaluated separately, each of which is exponentially small (see \secref{sec:measure}). For an observable defined for an unnormalized Choi state $\Vert \rho \rAngle$, it should be normalized by the norm $\lAngle \rho \Vert \rho \rAngle^{1/2}$, which is the square root of the purity of $\rho$ in the physical Hilbert space. In general, the purity of the decohered SPT state $\rho_\spt^D$ is exponentially small:
\begin{equation}
    \textrm{tr}(\rho^D_\spt \rho^D_\spt) \sim \exp(-N)
\end{equation}
where $N$ is the number of sites for any $p > 0$ in the decoherence model (c.f. \eqnref{eq:noise_basic}). 
Likewise, the numerator for the expression of an observable would decay exponentially with the system size. Therefore, although in principle we can define a quantity to characterize an ordering in the doubled Hilbert space, it only characterizes the presence of some nontrivial quantum entanglements whose verification is exponentially hard. A similar situation occurs in the famous black hole information paradox: the reconstruction of a quantum state fallen into an \emph{old} blackhole based on the Hawking radiation is theoretically possible, but it would require an exponential computational complexity in general~\cite{Hayden_2007, Yoshida2017, You1803.10425}.

\section{Experimentally probing the strange correlator} \label{sec:measure}

So far, we have discussed how strange correlators can
detect nontrivial SPT features even under decoherence.
Naively, strange correlators are just theoretical tools since we cannot sandwich two different states
in actual experiments. However, in this section, we will show that with the help of classical computations,
one can experimentally probe nontrivial signatures of
the strange correlators through measurements, so-called
computationally assisted observables (CAO)~\cite{Lee2022, Garratt2022}.

\subsection{Strange correlator against a generic trivial state}

To further proceed, we consider a strange correlator against a generic trivial disordered state  $\ket{\Omega_\bs}$, which is defined as the product state along the $X$-basis characterized by the bit-string $\bs = \{s_1,s_2,...,s_N\}$:
\begin{equation} \label{eq:generic_product}
    \ket{\Omega_\bs} \equiv \bigotimes_{n=1}^N \big[ Z^{(1-s_n)/2}|+\rangle_n \big],\quad  s_i \in \{-1,1\}
\end{equation}
which is invariant under the global symmetry. Essentially, the label $\bs$ specifies the symmetry charge configuration of the trivial state of interest.  Then, we define $C^\textrm{I}_\bs(r)=\tr(\rho_\text{spt}O(0)O(r)\rho_{0,\bs})/\tr(\rho_\text{spt}\rho_{0,\bs})$ as the type-I strange correlator computed against the $\bs$-labeled trivial state density matrix $\rho_{0,\bs} \equiv |\Omega_\bs \rangle \langle \Omega_\bs |$. We emphasize that for the strange correlator to be well-defined, the overlap between the SPT and trivial states $\tr(\rho_\spt \rho_{0,\bs})$ has to be non-zero. For that, it is necessary that they have components carrying the same total charge under global symmetry. 
In general, one may expect that this generalized strange correlator $C^\textrm{I}_\bs(r)$ still serves as a suitable probe for the SPT physics as the locally-charged background (specified by $\bs$) should not change the qualitative behaviors.

In the following, we compute strange correlators of pure $G_A \times G_B$ SPT states against generic trivial states for two exemplary cases with or without decoherence, where the decoherence model is given as \eqnref{eq:noise_basic}:

\vspace{5pt}

\noindent $(i)$ \emph{1d cluster $\mathbb{Z}_2^\todd \times  \mathbb{Z}_2^\teven$ SPT state}: 
For $|\Omega_\bs \rangle$ to have a non-zero overlap with the 1d cluster state, the trivial state should have even parity on both symmetries, i.e., $\prod_n s_{2n+1} = \prod_n s_{2n} = 1$. Then,
\begin{align} \label{eq:generic_SCI}
    C^\textrm{I}_{\todd,\bs}(2n) = \prod_{m=1}^n s_{2m},\quad C^\textrm{I}_{\teven,\bs}(2n) = \prod_{m=1}^n s_{2m-1}.
\end{align}
Under decoherence of $\mathbb{Z}_2^\teven$, $\rho_\spt^D$ would become an ensemble of configurations with different total $\mathbb{Z}^\teven_2$ charges, and the constraint $\prod_n s_{2n} = 1$ does not have to hold anymore for a trivial state. In this case, $C^\textrm{I}_{\teven,\bs}(2n)$ remains unchanged while $C^\textrm{I}_{\todd,\bs}(2n)$ becomes short-ranged as in \eqnref{eq:strange_typeI_result}. Even away from the stabilizer limit, the numerical results in \figref{fig:1d_numerics} has shown that the above sign structure persists, and the magnitude of $C^\textrm{I}_{\todd,\bs}$ does not depend on the $\mathbb{Z}_2^\teven$-charge configuration of the trivial state $\bs_\teven = \{s_2,s_4,...\}$. However, the magnitude of the $C^\textrm{I}_{\todd,\bs}$ can change depending on $\bs_\todd = \{s_1, s_3, ... \}$.

\vspace{5pt}
\noindent $(ii)$ \emph{2$d$ cluster $\mathbb{Z}^{\textrm{(0)}}_2 \times  \mathbb{Z}_2^{\textrm{(1)}}$ SPT state}: 
For $|\Omega_\bs \rangle$ to have a non-zero overlap with the 2d cluster state, there are two necessary conditions: $(i)$ $\prod_v s_v = 1$ (trivial 0-form global charge), and $(ii)$ $B_p \equiv \prod_{e \in p} s_e = 1$ for all plaquette $p$ (trivial 1-form charge configurations). Then, an ordinary correlation $Z_v Z_{v'}$ for $G_A = \mathbb{Z}^{\textrm{(0)}}_2$ and a Wilson loop operator $\prod_{e \in \rd \cA} Z_e$ for $G_B = \mathbb{Z}^{\textrm{(1)}}_2$ are given as follows
\begin{align}
    C^\textrm{I}_{A,\bs}(\rd l) = \prod_{e \in l} s_{e},\quad 
    C^\textrm{I}_{B,\bs}(\rd \cA) = \prod_{v \in \cA} s_{v}.
\end{align}
where $l$ is an open string connecting two vertices $v$ and $v'$, i.e., $\rd l = \{v,v'\}$. 

Under decoherence of $G_A$ (0-form), the above sign structure for $C^\textrm{I}_B$ remains robust, while the magnitude can change depending on the decoherence strength as in \eqnref{eq:1form_strangeIB}. In this case, $C^\textrm{I,II}_A$ is unaffected. Under decoherence of $G_B$ (1-form), however, $C^\textrm{I}_A$ gets modified in an interesting way. It is straightforward to show that under decoherence,
\begin{align} \label{eq:RBIM_Ising}
    C^\textrm{I}_{A,\bs} = \langle Z_{v} Z_{v'} \rangle_\beta^{\textrm{RBIM}(\bs)}
\end{align}
which is the correlation function of the random bond Ising model (RBIM) at the inverse temperature $\beta = \tanh^{-1}(1-2p)$ for the bond configuration $\bs$. 
As the correlation of the RBIM, the magnitude of $C^\textrm{I}_{A,\bs}$ only depends on the bond frustration, which corresponds to the 1-form charge configuration $\{ B_p \}$ of the trivial state $|\Omega_\bs \rangle$. 
Interestingly, unlike the 1d cluster state case, this implies that the type-I strange correlator of the mixed state $\rho^D_\spt$ behaves differently depending on the trivial state we computed against. 
For example, when $s_e = 1$, $C^\textrm{I}_{A,\bs}$ behaves as the correlation of the ferromagnetic Ising model. If we choose $s_e$ randomly such that $B_p = +1$ for half the plaquettes and $B_p = -1$ for the other half, $C^\textrm{I}_{A,\bs}$ behaves as the correlation of the maximally frustrated RBIM, decaying exponentially in distance for any $p$. 
This raises the question of which strange correlator is appropriate for one to use to diagnose the nontrivial behavior of the underlying SPT physics.
In fact, a proper way is to take a weighted average of all type-I strange correlators, whose weight is determined by the overlap between the trivial state from a given density matrix, $P_\bs = \langle \Omega_\bs | \rho_\spt^D | \Omega_\bs \rangle$. In this formalism, for small $p$ the weight for the latter scenario is extremely small, and the \emph{average} behavior of the type-I strange correlator would still be long-ranged. We will examine the implication of this in the next section.

\subsection{With post-selection}

With measurements and post-selections, it is possible to experimentally extract ($i$) the quantity very similar to the type-I strange correlator and ($ii$) the type-II strange correlator. The idea is that measurements and post-selections allow one to project the SPT state into a certain trivial product state to some degree, a step essential to compute the strange correlator. 
To illustrate the idea, consider a $1d$ cluster state discussed in the previous section. 
Assume that we want to understand the behavior of the type-I strange correlators for the $\mathbb{Z}_2^\todd$ symmetry. 
As the type-I strange correlator is not a physical quantity that can be directly measured, we instead define the \emph{marginalized} type-I strange correlator:
\begin{equation} \label{eq:marginalized_strangeI}
\begin{split}
    \tilde{C}^\textrm{I}_{\todd, \bs^\textrm{e}}(2n) & \equiv  \frac{\textrm{tr}(\rho^D_\spt Z_1 Z_{2n+1} \cP_{\bs^\textrm{e}} )}{\textrm{tr}(\rho^D_\spt \cP_{\bs^\textrm{e}}  )}  \\
   & = \langle Z_1 Z_{2n+1} \rangle_{\cP^\textrm{e}_{\bs^\textrm{e}} \rho^D_\spt}
\end{split}
\end{equation}
where $\cP_{\bs^\textrm{e}}$  is the projector onto a specific $\mathbb{Z}_2^\teven$-charge configuration defined as 
\begin{align} \label{eq:projector2}
    \cP_{\bs^\textrm{e}} \equiv  \prod_{m=1}^{N} \frac{1 + s_{2m} X_{2m}}{2}
\end{align}
with the bit-string ${\bs^\textrm{e}}=(s_2,s_4,...,s_{2N})$. We remark that the form of the marginalized type-I strange correlator is very similar to that of the type-I strange correlator. The crucial difference is that instead of evaluating against a trivial state density matrix $\rho_{0,\bs}$, which is essentially the projector onto a specific $\mathbb{Z}_2^\todd \times \mathbb{Z}_2^\teven$ charge configuration labeled by $\bs$, here we evaluate against the projector that only specifies the $\mathbb{Z}_2^\teven$ charge configuration. Thus, in the computation of $\tilde{C}^\textrm{I}_{\todd, \bs^\be}$, the $\mathbb{Z}_2^\todd$ charge configuration is \emph{marginalized}.

One may immediately notice that the marginalized type-I strange correlator is the correlation function of $Z$ operators for the \emph{projected} mixed state $\cP_{\bs^\textrm{e}} \rho^D_\spt \cP_{\bs^\textrm{e}}$. 
The projected mixed state can be obtained by measuring all the even sites in the $X$-basis and \emph{post-selecting} the resulting state with measurement outcomes equal to $\bs^\te$. 
As the correlation function of the physical mixed state, the marginalized type-I strange correlator is experimentally measurable. 
Also, as the name suggests, it is related to the type-I strange correlators. To show this, we define $\cP_{\bs^\oo}$ as the projector onto a specific $\mathbb{Z}_2^\todd$ configuration labeled by $\bs^\oo = (s_1,s_3,...,s_{2N-1})$. Together, $\cP_{\bs^\oo}$ and $\cP_{\bs^\te}$ project onto a configuration where all symmetry charges are specified, i.e., a trivial disordered state $|\Omega_{\bs}\rangle$ with $\bs = (\bs^\oo, \bs^\te)$.  Therefore, by inserting a completeness relation $\mathbb{I} = \sum_{\bs^\oo} \cP_{\bs^\textrm{o}}$ in \eqnref{eq:marginalized_strangeI}, we can show that
\begin{equation} \label{eq:tilde_strange_typeI}
\begin{split}
    \tilde{C}^\textrm{I}_{\todd,\bs^\textrm{e}}(2n) &= \sum_{\bs^\textrm{o}}  \frac{\textrm{tr}\big(\rho^D_\spt \cP_{\bs^\textrm{e}}  \cP_{\bs^\textrm{o}} \big)}{\textrm{tr}\big(\rho^D_\spt \cP_{\bs^\textrm{e}} \big)} \\
    &\hspace{30pt} 
    \frac{ \textrm{tr}\big(\rho^D_\spt Z_1 Z_{2n+1} \cP_{\bs^\textrm{e}}  \cP_{\bs^\textrm{o}} \big)}{ \textrm{tr}\big(\rho^D_\spt \cP_{\bs^\textrm{e}}  \cP_{\bs^\textrm{o}} \big) }  \\
    &= \sum_{\bs^\textrm{o}} P(\bs^\textrm{o} | \bs^\textrm{e}) \, C^\textrm{I}_{\todd, (\bs^\oo, \bs^\te)} (2n),
\end{split}
\end{equation}
where $P(\bs^\textrm{o}|\bs^\textrm{e} )$ is the probability of observing measurement outcome $\bs^\textrm{o}$ on odd sites conditioned on $\bs^\textrm{e}$ on even sites for the SPT state $|\Psi \rangle$. 
Therefore, as a weighted average of the type-I strange correlators, the experimentally accessible quantity $\tilde{C}^\textrm{I}_{\todd, \bs^\te}(2n)$ would quantify the overall non-trivialness of the type-I strange correlator.
In the stabilizer limit, as we showed in \eqnref{eq:generic_SCI}, the type-I strange correlator $C^\textrm{I}_{\todd, \bs}$ is independent of the values of $\bs$ on odd sites. Accordingly, the marginalized type-I strange correlator $\tilde{C}^\textrm{I}_{\todd, \bs^\te}$ would be equal to ${C}^\textrm{I}_{\todd, \bs}$ if $\bs$ matches $\bs^\te$ on even sites.

The type-II strange correlator is a quantity that can be expressed as the ratio of two fidelities: $\textrm{tr}(\rho_\spt^D \rho_{0,\bs}^\vdagger)$ and $\textrm{tr}(\rho_\spt^D {\rho}_{0,\bs'}^\vdagger)$ where $\rho_{0,\bs'} \equiv Z_1 Z_{2n+1} \rho_{0,\bs} Z_{2n+1} Z_1$ is also the pure state density matrix where $\bs'$ differs from $\bs$ in signs for two sites:
\begin{align} \label{eq:typeII_strange}
    C^\textrm{II}_{\todd,\bs}(2n) &= \frac{\textrm{tr}\big(\rho^D_\spt Z_{1} Z_{2n+1} \rho^\vdagger_{0,\bs} Z_{2n+1} Z_1 \big)}{\textrm{tr}\big(\rho^D_\spt \rho^\vdagger_{0,\bs} \big)} \nonumber \\
    &=\frac{\textrm{tr}\big(\rho^D_\spt \rho^\vdagger_{0,\bs'} \big)}{\textrm{tr}\big(\rho^D_\spt \rho^\vdagger_{0,\bs}  \big)} = \frac{P(\bs')}{P(\bs\,)} 
\end{align}
Therefore, if we repeatedly measure the charge configuration $\bs$ of the mixed state and estimate the probability $P(\bs)$ and $P(\bs')$ for the measurement outcome to be $\bs$ ($\bs'$), we can calculate $C^\textrm{II}_{\teven,\bs}(2n)$. Without post-selection, this is an extremely challenging task since there are exponentially many possible outcomes $\bs$. However, assuming we can post-select, one can compare the relative frequency between $\bs'$ and $\bs$ and estimate the type-II strange correlator. 

Another possibility to perform the fidelity estimation is to use classical shadow tomography \cite{Huang2002.08953}. In this approach, the decohered SPT state $\rho_\text{spt}^D$ will be repeatedly prepared and then measured in the random Clifford basis. The measurement will collapse $\rho_\text{spt}^D$ to a random stabilizer state $\ket{\sigma}$ with probability $\bra{\sigma}\rho_\text{spt}^D\ket{\sigma}$. These post-measurement states $\ket{\sigma}$ will be recorded and post-processed on a classical computer. In particular, the fidelity $\textrm{tr}(\rho_\text{spt}^D\rho_{0,\bs})$ (both in the numerator and the denominator) in \eqnref{eq:typeII_strange} can be estimated as $\textrm{tr}(\rho_\text{spt}^D\rho_{0,\bs})=\mathbb{E}_{\ket{\sigma}}(2^{2N}+1)\bra{\sigma}\rho_{0,\bs}\ket{\sigma}-1$, where $\bra{\sigma}\rho_{0,\bs}\ket{\sigma}$ can be efficiently computed given that $\ket{\sigma}$ is a stabilizer state and we have the freedom to choose the trivial reference state $\rho_{0,\bs}$ to be a stabilizer state as well. Thus both the numerator and the denominator of \eqnref{eq:typeII_strange} can be obtained by fidelity estimation via classical shadow tomography. However, this is not more efficient than the post-selection method. Because both fidelities are exponentially small ($\sim 2^{-2N}$) with respect to the system size $2N$, such that the number of measurements needed to accurately determine both the numerator and the denominator is still exponentially large ($\sim 4^{2N}$) in order to control the estimation standard deviation below the level of $2^{-2N}$.

\subsection{Without post-selection} \label{sec:nonlocal}

Post-selecting specific measurement outcomes allow us to access a type-II strange correlator as well as a weighted average of type-I strange correlators. 
However, as the system size increases, the number of samples for post-selection increases exponentially with the system size, and the whole idea becomes experimentally impractical. Therefore, for this method to be feasible, it is essential to avoid the post-selection issue. 
\subsubsection{Without decoherence}

Before addressing this issue, it is important to understand that in the pure SPT state, a nontrivial behavior of the type-I strange correlator in the SPT phase is intimately tied to the presence of finite non-local SPT order parameters, which can be experimentally measured to diagnose the underlying SPT order. 
For example, the following \emph{string order parameter} in the 1d $\mathbb{Z}^\todd_2 \times \mathbb{Z}^\teven_2$ SPT phase allows one to distinguish an SPT and trivial phases~\cite{SOP1989, SOP2008}:
\begin{equation}
    S_{2a+1,2b+1} = Z_{2a+1} Z_{2b+1} \hspace{-3pt} \prod_{m={a+1}}^{b}  \hspace{-3pt} X_{2m},
\end{equation}
which takes a non-zero expectation value as $|a-b| \rightarrow \infty$ in the SPT phase. 
One way to interpret this quantity is the following: in the SPT phase, the application of the $\mathbb{Z}_2^\teven$ symmetry transformation on a partial region is equivalent to threading a symmetry defect (flux) at the boundaries of the region. For this SPT state, by definition, such a $\mathbb{Z}_2^\teven$ symmetry defect should induce a $\mathbb{Z}_2^\todd$ charge, which can be canceled by the application of the charge creation/annihilation operator. If we define $S$ as such a partial symmetry transformation combined with the charge creation/annihilation operators at the boundary, $S$ would behave non-trivially, while for a trivial disordered state, the defect does not induce any charge and $S$ would take a value exponentially decaying in its size.

For a given SPT state $\rho_\spt$, the expectation value of the string order parameter can be expressed in terms of strange correlators as the following:
\begin{align} \label{eq:string_order}
    \textrm{tr}\big( \rho_\spt S_{1,2n+1} \big) &= \sum_{\bs} P_\bs \Big[ \prod_{m=1}^{n} s_{2m} \Big]\, C^\textrm{I}_{\todd, \bs}(2n)  \nonumber \\
    &= \sum_{\bs^e} P_{\bs^e} \Big[ \prod_{m=1}^{n} s_{2m} \Big]\, \tilde{C}^\textrm{I}_{\todd, \bs^e}(2n) 
\end{align}
where we used the completeness of the product states $\sum_{\bs} \rho_{0,\bs} = \mathbb{I}$ for the derivation, and $P_\bs = \textrm{tr}( \rho_\spt \rho_{0,\bs} )$ is the probability to obtain the measurement outcome $\bs$ by measuring $\rho_\spt$ in $X$-basis, satisfying $\sum_\bs P_\bs = 1$. 
The expression implies that the string order parameter is a weighted average of \emph{signed} strange correlators. 
The implications of \eqnref{eq:string_order} can be summarized as the following:
\begin{itemize}[leftmargin=16pt]
    \item If the expectation value of the non-local parameter is non-trivial, the signed average of the type-I strange correlators should be non-trivial.
    \item If $(i)$ the type-I strange correlators are nontrivial and $(ii)$ their sign structure cancels the prefactor $\prod s_{2m}$, the non-local parameter $S$ must be non-trivial. However, even if the condition $(i)$ holds, the condition $(ii)$ may not hold under decoherence.
\end{itemize}
In the pure SPT state where the expectation value of the non-local order parameter is nontrivial, indeed the strange correlators are nontrivial and their signs are generally canceled by the term $\prod s_{2m}$. In other words,
\begin{equation}
    C^\textrm{I}_{\todd,\bs} = \prod_{m=1}^n s^\vdagger_{2m} C^\textrm{I}_{\todd,\bar{\bs}} \,\, \Rightarrow \,\, \expval{S_{1,2n+1}} = \tilde{C}^\textrm{I}_{\todd, \bar{s}^e}
\end{equation}
where $\bar{\bs}$ is defined in such a way that its even part $\bar{\bs}^e = 1$ and its odd part $\bar{\bs}^o = \bs^o$. Therefore, the string order parameter would be given as the marginalized type-I strange correlator discussed in the previous section. Since the string order parameter can be simply measured from the bulk wavefunction without any post-selection, it implies that in the pure SPT state measuring string order parameters allow one to learn about the type-I strange correlators.

To be more concrete, we outline the following experimental procedure to measure non-local order parameters:
for a given $\rho_\spt^D$, measure all even sites by $X$-basis and odd sites by $Z$-basis, where $\bs^\textrm{e} = (s_2, s_4,...)$ and $\bsigma^\textrm{o} = (\sigma_1, \sigma_3, ...)$ are measurement outcome bit-strings. Now, we repeat this procedure repeatedly and obtain a set of $M$ bit-strings $\{ \bs^{\textrm{e},(i)} \}_{i=1}^M$ and $\{ \bsigma^{\textrm{o},(i)} \}_{i=1}^M$. Then, 
\begin{align} \label{eq:exp_quantity_typeI}
    & \frac{1}{M} \sum_{i=1}^M \Big[ \prod_{m=1}^n s^{(i)}_{2m} \Big]\,\sigma^{(i)}_{1} \sigma^{(i)}_{2n+1} \rightarrow \tilde{C}^\textrm{I}_{o,\bar{\bs}}(2n)
\end{align}
whose statistical uncertainty reduces \emph{polynomially} in the number of measurement snapshots $M$. 
This is exactly equivalent to the measurement of the marginalized type-I strange correlator except we multiply by the additional factor $\prod s_{2m}$. In fact, the absence of post-selection is exactly compensated by the presence of this additional factor. Therefore, even without post-selection, we can identify a nontrivial behavior of the marginalized type-I strange correlator with polynomial sample complexity.

To generalize this idea to a higher dimension, consider an $G_A \times G_B$ SPT characterized by the mixed topological response between $G_A$ and $G_B$ symmetries; in other words, the domain wall (defect) of $G_B$ is decorated with $G_A$-SPT states~\cite{Chen2014}. 
In this case, the non-local order parameter $S^A$ associated with $G_A$-symmetry on the area $\cA$ is defined as the following \cite{Yoshida2016,2022arXiv221002485Z}:
\begin{align}
    S^A_g(\cA) \equiv O_g(\rd \cA)  \cdot U_g(\cA) 
\end{align}
where $g \in G_B$ is the element of the $G_B$ symmetry group, $O_g(\rd \cA)$ is the operator that creates a $G_A$-SPT state along the boundary $\rd \cA$ decorating the $g$-domain wall, and $U_g(\cA)$ is the partial $G_B$ symmetry transformation on the region $\cA$.
By definition, this non-local order parameter would exhibit a perimeter law in the SPT phase, i.e., it decays exponentially with its perimeter $| \rd \cA |$:
\begin{equation}
    \langle S_g^A(\cA) \rangle \propto e^{-a | \rd \cA | }
\end{equation}
When $G_B$ is a $n$-form symmetry, the region of support $\cA$ is $(d\,{-}\,n)$-dimensional. Accordingly, its boundary is $(d\,{-}\,n\,{-}\,1)$-dimensional. Therefore, if $G_B$ is a $(d\,{-}\,1)$-form symmetry, there always exist a non-local order parameter for $G_B$ which takes a finite value in the SPT phase. On the other hand, if $G_B$ is $(d\,{-}\,2)$-form symmetry, then it has support on a 2-dimensional manifold, whose boundary is 1-dimensional. Accordingly, in the SPT phase, the associated non-local order parameter would decay exponentially with its boundary length; however, for the trivial state, this order parameter would decay exponentially with its area, so one can distinguish an SPT state from a trivial disordered state.


By inserting a completeness relation in the $G_B$-charge basis $\mathbb{I} = \sum_{\bs^B} \cP_{\bs^B}$, where $\bs^B$ is the string specifying a $G_B$-charge configuration, the non-local order parameter expectation value can be expressed as
\begin{align} \label{eq:SC_NLO}
    \textrm{tr}(   \rho_\spt^D S_g ) &= \sum_{\bs^B} P_{\bs^B} \, c(U_g|\bs^B) \, \tilde{C}^\textrm{I}_{A,\bs^B}( O_g )
\end{align}
where each term is defined as
\begin{align}
    P_{\bs^B} &\equiv \tr(\cP_{\bs^B} \rho_\spt^D ) \nonumber \\
    c(U_g|\bs^B) & \equiv  \bra{\Omega_{\bs^B}} U_g(\cA)\ket{\Omega_{\bs^B}}, \nonumber \\
    \tilde{C}^\textrm{I}_{A, \bs^B}(O_g) &\equiv \sum_{\bs^A} P(\bs^A| \bs^B) \, C^\textrm{I}_{A,(\bs^A,\bs^B)}(O_g).
\end{align}
$P_{\bs^B}$ is the probability of obtaining a measurement outcome $\bs^B$ on $\rho_\spt^D$, and $c(U_g|\bs^B)$ is the phase factor obtained by applying $U_g(\cA)$ on the product state labeled by $\bs^B$, and $\tilde{C}^\textrm{I}_{A,\bs^B}( O_g )$ is the marginalized type-I strange correlator.
Therefore, for these SPT states the points made earlier for the 1d $\mathbb{Z}_2 \times \mathbb{Z}_2$ SPT state immediately extend.

\subsubsection{With decoherence}

Despite the connection between the strange correlator and type-I strange correlator (c.f. \eqnref{eq:SC_NLO}), nontrivial behaviors of the type-I strange correlator under decoherence do not imply that the non-local order parameters are also nontrivial under decoherence.
As examined in \eqnref{eq:2dSPT_NLO_decoherence}, the non-local order parameter immediately becomes short-ranged under decoherence while the type-I strange correlator remains nontrivial for a weak enough decoherence strength. The reason is that $c(U|\bs^B)$ does not cancel the sign of $\tilde{C}^\textrm{I}_{A,\bs^B}$, and the weighted average of the marginalized strange correlators with unbalanced oscillating signs becomes short-ranged. 

However, this issue can be circumvented through some computational assistance~\cite{Lee2022}. 
Theoretically, the question can be rephrased as finding a decoding quantum channel $\cD$ such that if we evaluate a non-local order parameter of $\cD[\rho_\spt^D]$ instead of $\rho_\spt^D$, the quantity becomes non-trivial. Conceptually, this can be achieved by multiplying an additional $\bs^B$-dependent phase factor $d(S_g|\bs^B)$ such that it cancels the oscillating sign in the summation:
\begin{align} \label{eq:dec_relation}
    \textrm{tr}(   {\cD}[\rho_\spt^D] S_g ) &= \sum_{\bs^B} P_{\bs^B} \, c(U_g|\bs^B)  \, \tilde{C}^\textrm{I}_{A,\bs^B}( O_g ) \cdot  d(S_g|\bs^B).
\end{align}
If $c \cdot \tilde{C}^\textrm{I} \cdot d \geq 0$ for the majority of possible outcomes $\bs^B$, the above decoded observable can be non-trivial. In principle, if $\tilde{C}$ is nontrivial for each $\bs^B$, there must exist $d$ such that the above summation becomes non-trivial.

Experimentally, the procedure is as follows. First, we measure all $G_B$-charges. Then, based on the measurement outcome $\bs^B$, we can predict or \emph{decode} the sign for the corresponding marginalized type-I strange correlator, and apply a relevant transformation on the remaining qubits. This step is captured by the decoding quantum channel $\cD_{\bs^B}$ acting on $G_A$-charges~\footnote{Although it sounds like we apply unitary transformations, this procedure can be stochastic, implying that the decoding step is captured by the quantum channel.}.  
The combined action of measurement and decoding can be described by the following quantum channel:
\begin{equation}
    \cD\big[ \rho_\spt^D \big] \equiv \sum_{\bs^B} \cD_{\bs^B} \big[ {\cP}_{\bs^B} \rho_\spt^D {\cP}_{\bs^B} \big].
\end{equation}
where $\cP_{\bs^B}$ is the projector onto the measurement outcomes $\bs^B$ on the $G_B$-charges. In this representation,
\begin{equation}
    d(S_g|\bs^B) \equiv \frac{\tilde{C}^\textrm{I}_{A,\bs^B}( \cD_{\bs^B}[O_g] )}{\tilde{C}^\textrm{I}_{A,\bs^B}( O_g )}
\end{equation}
where we used the self-adjointness of the decoding channel $\cD_{\bs^B}$ which only acts on $G_A$-charge associated with the $O_g$ operator. Finally, we measure $O_g$ for this state projected into $\bs^B$, which is equivalent to measuring the marginalized type-I strange correlator as in \eqnref{eq:marginalized_strangeI}. Let $\bsigma^A$ be the label for the measurement outcome of $O_g$. One can repeat this procedure, and then use the sequence of $\{ (\bsigma^{A,(i)},\bs^{B,(i)})\,|\, i=1,2,... \}$ to estimate the \emph{computationally assisted observable}.


Our understanding here can be summarized by the following inequality, which can be obtained from \eqnref{eq:dec_relation}: 
\begin{align} \label{eq:inequality}
    \Big| \, \textrm{tr}(   {\cD}[\rho_\spt^D] S_g ) \Big| \leq   \sum_{\bs^B} P_{\bs^B} \, \Big| \tilde{C}^\textrm{I}_{A,\bs^B}( O ) \Big|. 
\end{align}
This inequality has important consequences:
\begin{itemize}
    \item Consider the case where $\cD$ is an identity quantum channel, i.e., do-nothing operation, the LHS corresponds to a simple measurement of nonlocal (string) order parameter $S_g$ in the decohered state. Then, a nonlocal SPT order parameter serves as a lower bound of the marginalized type-I strange correlator. For example, if $S_g$ is nontrivial (long-range correlated), the marginalized type-I strange correlator must be also nontrivial. However, the opposite does not hold; even when $S_g$ is trivial (short-range correlated), the marginalized type-I strange correlator can be nontrivial.
    \item Consider the case where $\cD$ is an efficient decoding quantum channel, which allows one to extract a nontrivial quantity in LHS. Then, this provides a theoretical guarantee that the underlying mixed state is a nontrivial SPT ensemble. 
    \item However, if the type-I strange correlators are already short-ranged, the RHS is trivial. Then, no matter how well we choose $\cD$, the LHS must be trivial. Therefore, the behavior of the type-I strange correlator imposes a fundamental bound on topological information we can efficiently extract from a decohered mixed state. 
\end{itemize}

\subsection{Stability of measurement-assisted state preparation under decoherence}

It has been pointed out that by measuring SPT (short-range entangled) states followed by appropriate feedback, one can realize long-range entangled quantum states, from GHZ states to non-abelian topological orders, and even fractons~\cite{1Dcluster_GHZ, 2Dcluster, 2Dcluster_toric, 3dCluster_fracton1, 3dCluster_fracton2, Stephen2017, Raussendorf2019, NatMeasurement, NatRydberg,CSScode, CSScode2, ClusterCSS, Lu2022, Lee2022, Zhu2022, nonAbelianNat, nonAbelianNat2}. In the measurement-assisted state preparations, one prepares an SPT state with two symmetries $G_A$ and $G_B$, and charges of one symmetry are measured, say $G_B$~\cite{NatMeasurement}. Then, the resulting state has a nontrivial correlation structure for $G_A$-charged operators; note that this correlation function is exactly equivalent to our definition of the marginalized type-I strange correlators in \eqnref{eq:tilde_strange_typeI}. Finally, one applies appropriate feedback based on the measurement outcomes of $G_B$-charges so that this nontrivial correlation of $G_A$-charged operators is transformed into a quantity independent of the measurement outcomes. The last step corresponds to finding a decoding quantum channel $\cD$ in our language. If the last step fails, the measurement-assisted state preparation fails as well.

Therefore, \eqnref{eq:inequality} provides a fundamental bound on what can be achieved in the context of measurement-assisted state preparations under decoherence. Since our inequality imposes a fundamental bound on the last step mentioned above, we get the following theorem:

\vspace{5pt}
{\noindent {\bf Theorem} Under decoherence, an SPT state can be diagnosed efficiently in the sample size only if the non-trivial marginalized type-I strange correlator exists.   
Furthermore, a scheme to prepare a long-range entangled state by measuring an SPT state can exist only if the corresponding marginalized type-I strange correlator is nontrivial. 
}
\vspace{5pt}

As we have examined for the 2d cluster state SPT, the type-I strange correlator is generally nontrivial under weak decoherence of 1-form symmetry, and its marginalized version can be shown to be nontrivial for small $p$. Then, the theorem implies that there must exist a decoding protocol to capture this quantity in a sample efficient manner, and also to facilitate the measurement-assisted state preparation. In Ref.~\cite{Lee2022}, this was shown explicitly where the decoding transformation $\cD_{\bs^B}$ corresponds to spin-flips on the vertex qubits based on the measurement outcomes $\bs^B$ on the edges (rotation in $Z$-axis followed by measurement in $X$-basis is equivalent to having decoherence by a dephasing channel). We remark that the theorem does not provide a decoding protocol, and finding an optimal decoding transformation can be computationally challenging. However, several cases exist with sub-optimal yet computationally efficient decoding solutions. Indeed, Ref.~\cite{Lee2022} also demonstrates a sample- and computation-efficient protocol for the 2d SPT state where the 1-form symmetry is decohered.

\section{Conclusion and Outlook}

In this work, we studied symmetry-protected topological ensembles described by density matrices with nontrivial topological features. We proposed several quantities that can be used to diagnose nontrivial topological features of \emph{mixed state} density matrices, namely the type-I and type-II strange correlators. Using these strange correlators, we have shown that an SPT state under \emph{selective} decoherence can still persist. We also introduced a ``doubled Hilbert space'' formalism based on the Choi-Jamiołkowski isomorphism, in which the mixed state is mapped to a pure state, and this doubled formalism gives us a general definition of SPT ensembles, as well as a unified picture of both type-I and type-II strange correlators. Under selective decoherence, the mapping provides a natural explanation for the robustness of the type-II strange correlator as a signature of the pure SPT state in the doubled Hilbert space under a shallow-depth symmetric quantum circuit. Finally, we interpreted the type-I and type-II strange correlators as indicators of the information-theoretic phases, quantifying the presence of nontrivial topological information in the mixed-state density matrix that may not be readily accessible in the experiments.

A natural yet speculative direction one may ask is the full implication of the Choi-Jamiolski isomorphism in understanding mixed state properties. The mapping allows one to understand nontrivial properties of the density matrix as physical properties of the corresponding Choi state in the doubled Hilbert space. 
Accordingly, one can classify density matrices based on the possible quantum phases of Choi states constrained by Hermiticity, positive semi-definiteness, and average symmetry. This may give an interesting way to quantify the information content of the density matrices.

Many subjects remain unexplored along the direction of decohered quantum states of matter. As we have mentioned, we will leave a more systematic study of decohered fermionic TIs, and TSCs to the future. For each element of the ten-fold way classification, there can be various types of decoherence channels, which likely leads to very rich physics. As we have seen in the paper, decoherence may be mapped to an interaction between the two boundaries of the SPTs, hence decoherence may lead to various phenomena such as symmetric mass generation~\cite{Wang1307.7480,wen2013,Slagle1409.7401,Ayyar1410.6474,you2014,yougrand,Catterall1510.04153,You1705.09313,You1711.00863,Xu2103.15865, Tong2104.03997, Wang2204.14271}, and decoherence collapsed classification of TIs and TSCs~\cite{Fidkowski0904.2197, Fidkowski1008.4138, Turner1008.4346, Ryu1202.4484, Qi1202.3983, Yao1202.5805, Gu1304.4569, Wang1401.1142, Metlitski1406.3032, You1409.0168, Cheng1501.01313, Yoshida1505.06598, Gu1512.04919, Song1609.07469, Queiroz1601.01596, Witten1605.02391, Wang1703.10937, Kapustin1701.08264}. When the classification of TIs and TSCs is collapsed due to interaction, it was shown that the original topological transition between the topological insulator and the trivial state may become an ``unnecessary transition''~\cite{bisenthil,jianxune}, namely this transition is still a stable fixed point in the parameter space, but two sides of the transition can still be connected adiabatically. We expect that the unnecessary transition may also occur due to decoherence in the doubled Hilbert space.

Another big class of quantum states of matter is the gapless quantum states, including quantum critical points. The effect of weak measurement on $(1+1)d$ conformal field theory (CFT) was studied in Ref.~\cite{Garratt2022}, and it was shown that the effect of weak measurement can be mapped to the boundary operators of the CFT. In the past few years, the boundary of $(2+1)d$ quantum critical points have attracted a lot of attention from both theoretical and numerical community~\cite{groverashvin,zhanglong1,zhanglong2,wessel1,wessel2,wessel3,xuboundary1,xuboundary2,max1,max2,max3,shang,toldin1,wessel2,zhanglong3}. The problem becomes particularly interesting when the $(2+1)d$ bulk was an SPT state driven to a critical point, hence two different boundary effects were at play. Quantum critical points under decoherence will be another direction worth exploring.

\begin{acknowledgments}
We thank Ehud Altman, Soonwon Choi, Matthew P. A. Fisher, Sam Garrett, Chao-Ming Jian, Joel Moore, Xiao-Liang Qi, Shinsei Ryu, and Ryan Thorngren for inspiring discussions. We would like to thank the KITP program ``Quantum Many-Body Dynamics and Noisy Intermediate-Scale Quantum Systems'' where the collaboration of this work was initiated. J.Y.L is supported by the Gordon and Betty Moore Foundation under the grant GBMF8690 and by the National Science Foundation under the grant PHY-1748958. Y.Z.Y. is supported by a startup fund at UCSD and the NSF Grant No. DMR-2238360. C. X. is supported by the NSF Grant No. DMR-1920434, and the Simons Investigator program.

\emph{Note Added:} While finishing up this work, we became aware of an independent related work~\cite{Zhang2022S2210.17485}.  
\end{acknowledgments}

\bibliographystyle{quantum}



\onecolumn
\appendix

\section{More example} \label{app:example}

In this section, we explicitly demonstrate the application of strange correlators beyond $\mathbb{Z}_2$ symmetries. As a concrete example, we show that a strange correlator detects $1d$ SPT with $\mathbb{Z}_3 \times \mathbb{Z}_3$ symmetries. Consider a system with local Hilbert space dimension $3$ with basis $\{ |0 \rangle, |1 \rangle, |2 \rangle \}$. Let $\omega = e^{2\pi i /3}$. Then, $Z |n \rangle = \omega^n |n \rangle$, and $X |n \rangle = | (n+1) \mod 3 \rangle$. Furthermore, $ZX = \omega XZ$. The system has $\mathbb{Z}_3^\textrm{even} \times \mathbb{Z}_3^\textrm{odd}$ symmetries generated by $g_\textrm{even} = \prod_i X_{2i}$ and $g_\textrm{odd} = \prod_i X_{2i+1}$. Since the second group cohomology $H^2(\mathbb{Z}_3 \times \mathbb{Z}_3, \U(1)) = \mathbb{Z}_3$, there are three possible SPT phases, denoted by the label $m=0,1,2$ ($m=0$ corresponds to the trivial paramagnet). The Hamiltonian that realizes an SPT labeled by $m$ is given by~\cite{1dZnSPT}
\begin{align} \label{eq:Z3ham}
    H^{(m)} &= - \sum_i (X_{2i} (Z_{2i-1} Z_{2i+1}^\dagger)^m + \textrm{h.c.} ) \nonumber \\
    & \qquad - \sum_i (X_{2i+1} (Z^\dagger_{2i} Z_{2i+2})^m + \textrm{h.c.} )   
\end{align}
It is straightforward to show that all terms are commuting. Accordingly, the ground state $|\Psi_p \rangle$ is characterized by the unit expectation values of the following non-local (string) order parameters for all $r$:
\begin{align}
    S^m_{g,\textrm{even}}(2r) &:= (Z_0 Z^\dagger_{2r})^m (\prod_{i=0}^{r-1} X_{2i+1} )  \nonumber \\
    S^m_{g,\textrm{odd}}(2r) &:=(Z_1 Z^\dagger_{2r+1})^m (\prod_{i=1}^{r} X_{2i} )  
\end{align}
where $S_\textrm{even}$ measures $\mathbb{Z}_3^\textrm{even}$ charge response under the $\mathbb{Z}_3^\textrm{odd}$ flux insertion (and opposite for $\mathbb{Z}_3^\textrm{odd}$).

For the ground state of this fixed-point Hamiltonian \eqnref{eq:Z3ham}, we can consider a decoherence channel applied to it. If the decoherence channel respects the doubled symmetry, then it trivially preserves the expectation values $\langle S_g \rangle = 1$ for all $r$. For more nontrivial examples, we can consider a decoherence channel that is \emph{weakly} symmetric in $\mathbb{Z}_3^\textrm{even}$ (in the doubled space, it breaks $\mathbb{Z}_{3,u}^\textrm{even} \times \mathbb{Z}_{3,l}^\textrm{even}$ down to $\mathbb{Z}_{3,\textrm{diag}}^\textrm{even}$):
\begin{align} 
    &\mathcal{E}_{p,i}: \rho \rightarrow (1-p) \rho  + \frac{p}{2} (Z_i \rho Z^\dagger_i + Z^\dagger_i \rho Z_i), \quad \cE_p \equiv \prod_i \cE_{p,2i}
\end{align}
where $\rho^D_\spt = \cE_p[|\Psi_p \rangle \langle \Psi_p | ]$. How would strange correlators behave under this decoherence channel? To discuss this, we first define a trivial state
\begin{align}
    |\Omega \rangle := \prod_n \qty( \frac{|0\rangle + |1\rangle + |2\rangle }{\sqrt{3}} ), \quad \rho_0 = |\Omega \rangle \langle \Omega |
\end{align}
With this, for $\mathbb{Z}_3^\textrm{even}$ we define
\begin{align} 
    C_{\textrm{even}}^\textrm{I}(2r) &= \frac{\textrm{tr}( \rho^D_\spt (Z_0 Z^\dagger_{2r})^p \rho_0)}{ \textrm{tr}( \rho^D_\spt \rho_0) }   \nonumber \\
    C_{\textrm{even}}^\textrm{II}(2r) &= \frac{\textrm{tr}( \rho^D_\spt (Z_0 Z^\dagger_{2r})^p \rho_0 (Z_0 Z^\dagger_{2r})^p )}{ \textrm{tr}( \rho^D_\spt \rho_0) }.
\end{align}
Strange correlators for $\mathbb{Z}_3^\textrm{odd}$ can be also defined similarly. In order to facilitate the calculation, it is convenient to express $\rho_0$ as the following:
\begin{align}
    \rho_0 = \prod_i \qty( \frac{1 + X + X^\dagger}{3})
\end{align}
because $(X+X^\dagger)$ takes eigenvalues $2,-1,-1$. Then, we can show that
\begin{align}
    \cE_p[\rho_0] = \prod_i \qty( \frac{1 + (1-3p/2)(X + X^\dagger)}{3})
\end{align}
Using this relation, let us calculate strange correlators for a nontrivial SPT labeled by $m=0$:
\begin{align}
    C^\textrm{I}_{\textrm{odd}}(2r) &= e^{-r/\xi}, \quad \xi = 1/\ln(1/(1-3p/2)) \nonumber \\
    C^\textrm{II}_{\textrm{odd}}(2r) &= 1,
\end{align}
and $C^\textrm{I}_{\textrm{even}}(2r) = C^\textrm{II}_{\textrm{even}}(2r) = 1$. All these results are compatible with the decohered $\mathbb{Z}_2 \times \mathbb{Z}_2$ SPT phase discussed in \secref{sec:1dcluster}. The generalization into $\mathbb{Z}_n \times \mathbb{Z}_n$ can be done in an exact same fashion, where one can simply take $\omega = e^{2\pi i/n}$ and $m \in \{0,1,...,n-1\}$ while maintaining the Hamiltonian structure in \eqnref{eq:Z3ham}.

\section{Lemma} \label{app:note}

In this appendix, we demonstrate the following lemma that is used to facilitate calculations in the main text:

\vspace{5pt}

\noindent {\bf Lemma} Consider a $d$-dimensional cluster SPT state exhibiting a mixed topological response between two symmetries $G^\vdagger_A$ and $G^\vdagger_B$, where $G_A\,{=}\,G^{(n)}_1$ and $G_B\,{=}\,G^{(d-n-1)}_2$. 
Here, the superscripts represent that they are $n$ and $(d\,{-}\,n\,{-}\,1)$ form symmetries. Also, $G_A$ and $G_B$ act on different sublattices, say $L_A$ and $L_B$. Then, the expectation value of an operator $O$ acting only on $L_A$ or $L_B$ does not vanish only if $O$ contains a symmetry action on the corresponding sublattice.  

\vspace{5pt}

{\noindent \emph{Proof}: }  A $d$-dimensional cluster state with two symmetries $G_1$ and $G_2$ is defined by the following Hamiltonian consisting of two types of local terms:
\begin{align}
    H &= - \sum_{a \in L_1} h_{1,a} - \sum_{b \in L_2} h_{2,b}, \nonumber \\
    h_{1,a} &= {\cal O}^\textrm{d.w.}_{1,a} \prod_{b \in \rd a} {\cal O}^\textrm{charge}_{2,b}  \nonumber \\
    h_{2,b} &= {\cal O}^\textrm{d.w.}_{2,b} \prod_{a \in \dd b} {\cal O}^\textrm{charge}_{1,a}
\end{align}
where ${\cal O}_i^\textrm{charge}$ (${\cal O}_i^\textrm{d.w.}$) creates the charge (domain wall) of the symmetry $G_i$. As a cluster state Hamiltonian, $[h_{i,a}, h_{j,b}] = 0$. In fact, the result is generic for any SPT state with a stabilizer structure.

Now, consider an operator $O$ defined on the sublattice $L_1$. In general, $O$ can be decomposed into terms that carry quantum numbers under the set of stabilizers $\{ h_{1,a} \}$ and $\{ h_{2,b} \}$. However, note that any term that carries a nontrivial quantum number would have a vanishing expectation value since this stabilizer can be pulled out from the state and be absorbed back freely for the groundstate. In other words, the only term $O_A$ that can take a non-zero expectation value satisfies the following condition:
\begin{equation}
    \forall i,\,\,\,  h_{1,i} O_A h_{1,i} =  O_A \quad  \textrm{ and } \quad \forall j,\,\,\,  h_{2,j} O_A h_{2,j} = O_A 
\end{equation}
In fact, this already implies that $O_A$ is in fact a symmetry of the system. Since $O_A$ is restricted to $L_1$, it means that $O \in G_1$. A similar argument can be made for $O$ defined on the sublatice $L_2$. This concludes the proof.

\section{Choi-Jamiołkowski isomorphism} \label{app:Choi}

For a given density matrix $\rho$ defined on the Hilbert space ${\cal H}$, the Choi state is defined in the doubled Hilbert space ${\cal H}_d = {\cal H}_u \otimes {\cal H}_l$ as the following:
\begin{align}
        \Vert \rho \rAngle & \equiv \frac{1}{\sqrt{ \textrm{Dim}[\rho] }} \sum_i |i \rangle_u \otimes \rho |i\rangle_l \nonumber \\
        & = \frac{1}{\sqrt{ \textrm{Dim}[\rho] }} \sum_f \lambda_j  | \psi^*_j \rangle_u \otimes | \psi_j \rangle_l
\end{align}
where $\rho  = \sum_j \lambda_j |\psi_j \rangle \langle \psi_j|$. The Choi state automatically respects the following symmetry:
\begin{align}
    \textrm{SWAP}^* \equiv {\cal C} \circ \textrm{SWAP}
\end{align}
where the SWAP symmetry exchanges ${\cal H}_u$ and ${\cal H}_l$, and ${\cal C}$ is the complex-conjugation symmetry. This corresponds to Hermitian conjugation in the matrix language.

Now, for any state $| \Psi \rangle$ defined on ${\cal H}_d$ to be a Choi state, it should have the following form of Schmidt decomposition between upper and lower Hilbert spaces ${\cal H}_u$ and ${\cal H}_l$:
\begin{align}
    | \Psi \rangle &= \sum_i \lambda_i | \psi^*_i \rangle_u \otimes | \psi_i \rangle_l
\end{align}
where $\lambda_i \geq 0$. Note that, if we reshape a Choi state into the matrix form, the SWAP$^*$ symmetry guarantees that it is Hermitian. Accordingly, the above condition simply means that the eigenvalues of this matrix form should be non-negative. This imposes a strong condition on valid Choi states.

Let $G$ be the symmetry of the original system. The Choi state would enjoy a doubled symmetry group $G_u \times G_l$. We remark that the symmetry representation of $g \in G$ in the upper Hilbert space ${\cal H}_u$ is defined as a complex conjugated version of the original representation, $U^*(g)$. Accordingly, 
\begin{align}
    |\Psi \rangle \mapsto (U^*(g_u) \otimes U(g_l)) |\Psi \rangle \quad \forall g_u \cdot g_l \in G_u \times G_l
\end{align}

In general, Choi states mapped from density matrices respect the symmetry group $G_d \times \mathbb{Z}^*_2$, where $G_d$ is the diagonal subgroup of $G_u \times G_d$ and $\mathbb{Z}^*_2$ is the SWAP$^*$ symmetry. On the other hand, Choi states mapped from pure state density matrices respect the larger symmetry group
\begin{align}
    (G_u \times G_l) \rtimes \mathbb{Z}^*_2
\end{align}
where the SWAP symmetry exchanges $G_u \leftrightarrow G_l$. 

\section{A review for strange correlator and the NLSM description of SPT phases} \label{app:NLSMreview}

In this section, we provide a self-contained presentation of ($i$) the strange correlator and $(ii)$ effective field theory description of SPT phases using the non-linear sigma model (NLSM), for {\it pure} quantum states. 

Following the discussion in Ref.~\cite{sptwf}, without loss of generality, the matrix element of a pure state density matrix $|\Psi\rangle\langle \Psi|$ between two configurations $|\phi(\vect{x})\rangle$ and $|\phi'(\vect{x})\rangle$ can be expressed as a Euclidean space-time path-integral: 
\begin{align}
    \langle \phi(\vect{x})|\Psi\rangle\langle \Psi| \phi'(\vect{x}) \rangle \sim \int D[\phi(\vect{x}, \tau)]  e^{ - \int_{-\infty}^{+\infty} d\tau \int d^dx  \ \cL^{\Psi}[\phi] }
\end{align}
where $\cL^{\Psi}[\phi]$ is the Euclidean space-time Lagrangian, whose corresponding Hamiltonian has the ground state $|\Psi\rangle$. 
We assume that all the states  $|\phi(\vect{x})\rangle$ form a complete and orthogonal basis of the Hilbert space, and $\phi$ is a local field that carries a nontrivial representation of the symmetry group of the system. In the path integral above, we fixed the boundary condition $\phi(\vect{x}, \tau = -\infty) = \phi(\vect{x})$, and $\phi(\vect{x}, \tau = +\infty) = \phi'(\vect{x})$. This expression should be valid for any configurations $|\phi(\vect{x})\rangle$ and $|\phi'(\vect{x})\rangle$. Using the path integral formalism, the state $|\Psi\rangle$ can then be represented as
\begin{align}
    |\Psi\rangle \sim  \int D[\phi(\vect{x}, \tau)] e^{ - \int_{-\infty}^{0} d\tau \int d^d x \cL^{\Psi} }   | \tilde{\phi}(\vect{x})\rangle, 
\end{align}
where $\tilde{\phi}(\vect{x}) = \phi(\vect{x}, - \infty)$.

Hence in the path-integral formalism, the strange correlator defined in \eqnref{eq:strange_basic} is simply the correlation function at Euclidean time $\tau = 0$, which is at the temporal interface between two Lagrangians $\cL^\Psi$ and $\cL^\Omega$ ($\cL^\Omega$ is the Langriangian corresponding to a trivial disordered state $|\Omega \rangle$). As originally explained in Ref.~\cite{wenspt,wenspt2}, the group cohomology classification and characterization of SPT phases can be viewed as classifying and constructing topologically nontrivial actions in the (triangularized) Euclidean space-time. Therefore, if $\cL^\Psi$ and $\cL^\Omega$ are Lagrangians for SPT and trivial phases respectively, the strange correlator must inherit the 't Hooft anomaly that arises from the bulk topological action, just like the fact that the spatial boundary of a SPT phase must encodes the information of the 't Hooft anomaly. 

Suppose the bulk SPT state has spatial dimension $d$ and space-time dimension $D\,{=}\,(d+1)$, then its spatial boundary and temporal boundary both have total space-time dimension $D\,{=}\,d$. The 't Hooft anomaly at a boundary means that, if we ``gauge'' all the symmetries of the system, integrating out the matter fields at the boundary should generate anomalous terms that are not completely gauge invariant. Then for $d\,{=}\,1$ and $2$, the 't Hooft anomaly excludes the possibility that all the correlation functions between operators carrying nontrivial representations of the symmetry groups are short-ranged at the boundary (both temporal and spatial). This is because if all the correlations are short-ranged, integrating out the matter fields at the boundary would only generate local gauge invariant response to the gauge fields. 

But some subtlety arises when $d\,{=}\,3$, $i.e.$ the bulk is a SPT phase with three spatial dimensions. In principle, the 't Hooft anomaly at the $d\,{=}\,2$ spatial boundary (space-time dimension $D\,{=}\,3$) can also manifest as an anomalous topological order, i.e. even if all the correlation functions of local operators are short-ranged, the boundary system with total space-time dimension $D\,{=}\,3$ can still be anomalous. Example of symmetric anomalous topological order at the two dimensional boundary of a three dimensional SPT can be found in Ref.~\cite{MaxTO,ChenTO,WangTO,BondersonTO}. This scenario can in principle also happen at the temporal boundary of a three dimensional SPT phase, hence in principle the strange correlator of a three dimensional SPT phase can be trivially short-ranged, although the authors are not aware of an explicit example of such. 

Let us now review the NLSM description of the SPT phases. An O(3) NLSM was first introduced to describe the continuum limit dynamics of spin chains~\cite{haldane1,haldane2}, and it has become well-known that in order to capture the topological features of the Haldane phase, the NLSM in the continuum limit must have a topological $\Theta-$term whose fixed point is at $\Theta\,{=}\,2\pi$~\cite{haldanetheta,haldane1988b,affleck} (for the derivation, see for instance textbooks~\cite{fradkinbook,sachdevbook}): \beqn \cS = \int dx d\tau \frac{1}{g_0} (\partial_\mu \vect{n})^2 + \frac{\ii \Theta}{4\pi} \epsilon_{abc} n^a \partial_x n^b \partial_\tau n^c, \eeqn where $\vect{n}$ is the three component vector with unit length whose physical meaning is simply the normalized N\'{e}el order parameter. 

For $\Theta\,{=}\,2\pi$, this topological term does not lead to any special dynamics in the $(1+1)d$ bulk space-time (the bulk is a disordered phase with $g_0$ flows to $+\infty$ under RG), but it leads to important topological imprint on the boundary~\cite{ng1994}. This is because the $\Theta-$term at $\Theta\,{=}\,2\pi$ does not lead to any nontrivial factor in the path-integral when (and only when) the space-time manifold is compact. Hence the contribution of the $\Theta-$term only depends on the configuration of $\vect{n}$ on the space-time boundaries. Furthermore, one can safely ignore $\vect{n}$ in the bulk since the bulk is gapped and disordered, and arrive at an effective theory at the boundary, with potentially more nontrivial physics. The $(0+1)d$ spatial boundary of the system acquires a Wess-Zumino-Witten (WZW) term which is the topological ``legacy'' from the bulk~\cite{ng1994}, in addition to the ordinary dynamics of a NLSM: \beqn \cS_b = \int d\tau \frac{1}{\tilde{g}} (\partial_\tau \vect{n})^2 + \frac{2\pi k \ii}{4\pi} \int_0^1 du \ \tilde{n}^a \partial_\tau \tilde{n}^b \partial_u \tilde{n}^c, \eeqn where $u$ is an extra parameter introduced for the WZW term, and $\tilde{n}^a (\tau, u = 0) = n^a(x)$, $\tilde{n}^a (\tau, u = 1) = (0, 0, 1)$. When $\Theta = 2\pi$ in the bulk, the WZW term has level $k = 1$. The $(0+1)d$ NLSM plus the WZW term describes a dangling spin-$S$ with $S = k/2$, which is precisely the boundary physics of the Haldane phase. 

After the discovery of SPT phases in general dimensions with group cohomology classification~\cite{wenspt,wenspt2}, it was proposed that the NLSM in higher dimensions can provide a physically intuitive and convenient way of describing and understanding the SPT phases. This approach has been explored in previous literature such as Ref.~\cite{levinsenthil,ashvinsenthil2012,xu3d,binlsm}. The example most relevant to the current paper, is the bosonic SPT phase in $(2+1)d$ with (for example) a $\U(1)\times \U(1)$ symmetry, which can be described by the following NLSM~\cite{levinsenthil}: 
\begin{align}\label{2+1dnlsm}
    \cS = \int d^2x d\tau \ \frac{1}{g_0} (\partial_\mu \vect{n})^2 + \frac{\ii 2\pi}{\Omega_3} \epsilon_{abcd}n^a \partial_x n^b \partial_y n^c \partial_\tau n^d, 
\end{align}
where $\vect{n}$ is a four-component unit vector field, and $\Omega_3$ is the volume of a three-dimensional unit sphere. This NLSM serves multiple purposes: 
\begin{enumerate}[leftmargin=13pt]
    \item When the $(2+1)d$ bulk is disordered ($g_0$ being greater than certain critical value), this NLSM reduces to a $(1+1)d$ NLSM at the spatial boundary, and becomes a $(1+1)d$ NLSM defined with a unit four component vector $\vect{n}$ plus a WZW term: 
    \begin{align}
        \cS_b &= \int dx d\tau \ \frac{1}{\tilde{g}} (\partial_\mu \vect{n})^2 \nonumber \\
        &\quad \, + \frac{\ii 2\pi k}{\Omega_3} \int_0^1 du \ \epsilon_{abcd} \tilde{n}^a \partial_x \tilde{n}^b \partial_\tau \tilde{n}^c \partial_u \tilde{n}^d. \label{1+1dnlsme}    
    \end{align}
    Again, $u\,{\in}\,(0, 1)$ is an extra parameter, and $\tilde{n}^a(x, \tau, u\,{=}\,0)\,{=}\,n^a(x, \tau)$, and $\tilde{n}(x, \tau, u\,{=}\,1) = (0,0,0,1)$. For more detailed discussion of the dimensional reduction of the bulk $\Theta-$term to the boundary WZW term, please refer to Ref.~\cite{xuludwig,liuwen}. The exact value of the coupling constant $\tilde{g}$ in \eqnref{1+1dnlsme} is not important, as it will flow under RG. The maximal symmetry that \eqnref{1+1dnlsme} can host is SO(4), and with the SO(4) symmetry $\tilde{g}$ will flow to a fixed point which corresponds to the SU(2)$_k$ CFT~\cite{witten1984}. When $\Theta = 2\pi$ in the bulk, the WZW term has level $k = 1$ at the boundary. When the SO(4) is broken down to $\U(1) \times \U(1)$, the theory can be studied through Abelian bosonization, and the scaling dimensions of the CFT is determined by the Luttinger parameter. 

    \item When we reduce the NLSM to the temporal boundary, we can derive the wave function for the SPT phase \eqnref{2dnlsm}, following the discussions in Ref.~\cite{sptwf,YouXu2013}. The topological $\Theta-$term in the $(2+1)d$ bulk reduces to a $(2+0)d$ WZW term in \eqnref{2dnlsm}, and the $(2+0)d$ WZW term becomes the weight of the superposition of the field configuration $|\vect{n}(\vect{x})\rangle$. 
    
    The strange correlator calculation in this formalism becomes mathematically equivalent to computing the ordinary correlation functions of the $(1+1)d$ CFT described by \eqnref{1+1dnlsme}, and one can employ the powerful methods such as Abelian bosonization to evaluate the strange correlators. 
    
    \item The NLSM also makes the ``decorated defect'' construction of the SPT phases explicit. The decorated defect construction of SPT phases is the physical picture behind the K\''{u}nneth formula of the cohomology classification~\cite{Chen2014}, implying that a higher dimensional SPT phase can be constructed by decorating a lower dimensional SPT phases on symmetry defects, followed by proliferating the defects. This decoration becomes manifest in the NLSMs such as \eqnref{2+1dnlsm}. For example, if one creates a vortex of $(n_1, n_2)$, the $\Theta-$term of \eqnref{2+1dnlsm} reduces to a $(0+1)d$ $\Theta-$term which describes a $(0+1)d$ SPT phase with a U(1) symmetry that rotates $(n_3, n_4)$. For more details of the decorated defect interpretation of the NLSM and topological terms in general dimensions, please refer to Ref.~\cite{ashvinsenthil2012,binlsm}.

\end{enumerate}

Here we would like to clarify that, the strange correlator not only can determine whether a given wave function of a quantum many-body system $|\Psi\rangle$ is a nontrivial SPT state or not, it can also discern which SPT state it is, when there are more than one class of SPT phases for a given symmetry and spatial dimension. For this purpose we just need to use a simple generalization of the strange correlator, \beqn C_i(r) = \frac{\langle \Psi | O(0) O(r) |\Omega_i \rangle}{\langle \Psi | \Omega_i \rangle}, \label{Ci} \eeqn where $|\Omega_i \rangle$ is the representative wave function of the $i-$th SPT phase for the given symmetry. Within all these strange correlators, only one is supposed to be trivial, i.e. the one that is within the same class as $|\Psi\rangle$. Each of the strange correlator described above corresponds to the correlation function at the imaginary temporal interface between phase $|\Psi\rangle$ and the $i-$th SPT phase. Let us assume that $|\Psi\rangle$ actually belongs to the same class as the representative wave function $|\Omega_n\rangle$, then the temporal interface is anomaly-free only when $i = n$, hence only $C_i(r)$ with $i = n$ can be trivial. 

Generalization of Eq.~\ref{Ci} to the type-I and type-II density matrix form is straightforward.

\end{document}